# Atomic Resolution Observations of Nanoparticle Surface Dynamics and Instabilities Enabled by Artificial Intelligence

Peter A. Crozier[1]*, Matan Leibovich[2], Piyush Haluai[1], Mai Tan[1], Andrew M. Thomas[3], Joshua Vincent[1], Sreyas Mohan[4], Adria Marcos Morales[4], Shreyas A. Kulkarni[4], David S. Matteson[5], Yifan Wang[1], and Carlos Fernandez-Granda[2,4]*

*1. School for Engineering of Matter, Transport & Energy, Arizona State University, Tempe, AZ*

*2. Courant Institute of Mathematical Sciences, New York University, New York, NY*

*3. Department of Statistics & Actuarial Science, University of Iowa, Iowa City, IA*

*4. Center for Data Science, New York University, New York, NY*

*5. Department of Statistics & Data Science, Cornell University, Ithaca, NY*

* Corresponding authors: crozier@asu.edu and cfgranda@cims.nyu.edu

Nanoparticle surface structural dynamics is believed to play a significant role in regulating functionalities such as diffusion, reactivity, and catalysis but the atomic-level processes are not well understood. Atomic resolution characterization of nanoparticle surface dynamics is challenging since it requires both high spatial and temporal resolution. Though ultrafast transmission electron microscopy (TEM) can achieve picosecond temporal resolution, it is limited to nanometer spatial resolution [1-3]. On the other hand, with the high readout rate of new electron detectors, conventional TEM has the potential to visualize atomic structure with millisecond time resolutions. However, the need to limit electron dose rates to reduce beam damage yields millisecond images that are dominated by noise, obscuring structural details. Here we show that a newly developed unsupervised denoising framework based on artificial intelligence enables observations of metal nanoparticle surfaces with time resolutions down to 10 ms at moderate electron dose. On this timescale, we find that many nanoparticle surfaces continuously transition between ordered and disordered configurations. The associated stress fields can penetrate below the surface leading to defect formation and destabilization making the entire nanoparticle fluxional. Combining this unsupervised denoiser with electron microscopy greatly improves spatio-temporal characterization capabilities, opening a new window for future exploration of atomic-level structural dynamics in materials.

The concept of fluxionality, where a system rapidly moves through different isomers, was first discussed for organometallic molecules in the 1950s, as summarized by Cotton [5]. In the early days of nanoscience, there was interest in fluxional behavior of nanoparticles due in part to observations performed on newly-developed atomic resolution electron microscopes [6]. However, older, less sensitive electron detector technology limited temporal resolutions to about 100 ms and required large electron dose rates (> $10^4$ e Å$^{-2}$ s$^{-1}$). With the continued development of gas and liquid cell TEM, catalysis has been a primary motivation for studying structure and functionality in nanoparticles [7-10]. However, to limit beam damage, the time scale for much of the reported atomic structural dynamics is often minutes [11, 12]. Recently time resolutions on the order of 10 ms have been reported, but they employed high electron dose rates (≥ $10^4$ e Å$^{-2}$ s$^{-1}$) [13, 14]. Here, we use the power of machine learning to reduce the electron dose rate by at least an order of magnitude (~ $10^3$ e Å$^{-2}$ s$^{-1}$) while achieving temporal resolutions of ~ 10 ms and spatial resolutions of 1 Å. This enables us to explore the challenging issue of surface dynamics in metal particles.

To address the image noise challenge, we propose a denoising framework based on artificial intelligence (AI), which enables recovery of atomic-resolution information from noisy images. AI models based on neural networks have achieved impressive results for natural images, but often require training datasets with ground-truth clean images [15, 16]. Simulating such a dataset is challenging, and often impossible when the goal of denoising is scientific discovery. Here we propose a fully unsupervised framework to train and evaluate AI-powered denoising models using exclusively real noisy data [17]. The framework enables recovery of atomic-resolution information from TEM data, improving the signal-to-noise ratio (SNR) by a factor of 40 at a spatial resolution of 1 Å and time resolution near 10 ms. This enhanced time resolution reveals that supposedly stable, low-energy nanoparticle surfaces can display highly active atom dynamics, triggering instabilities resulting in rapid structural fluctuations. The new spatiotemporal capability enabled by the proposed AI framework dramatically enhances our ability to explore surface dynamics and the evolution of metastable states in nanoparticles at the atomic level, offering new insights into their evolving structures.

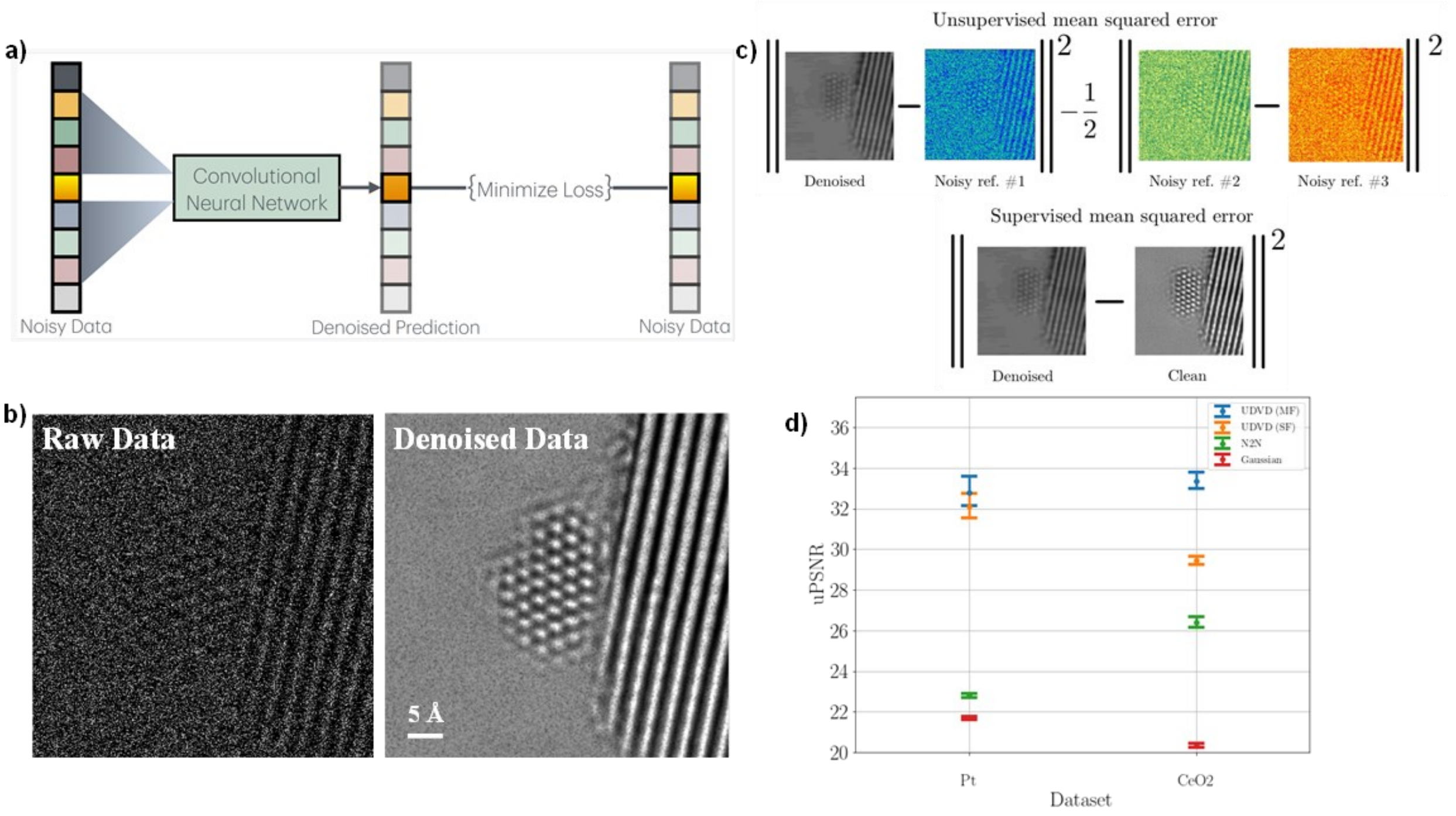

**Figure 1: a)** ***Unsupervised Deep Denoising:*** The proposed Unsupervised Deep Video Denoiser (UDVD ) learns to remove noise from noisy datasets without access to ground-truth clean images. This is achieved by training a deep convolutional neural network to estimate each noisy pixel from its spatio-temporal surrounding, but without using the noisy pixel itself. Since the noisy component of the pixel is unpredictable, the network learns to estimate the underlying clean signal. ***b) Example of Denoising Results:*** Images showing structure of Pt nanoparticle in a CO atmosphere at room temperature. Left is raw data (13ms exposure time), right same frame after UDVD denoising. ***c) Unsupervised Evaluation:*** In order to perform quantitative evaluation of unsupervised denoisers, we propose a metric called unsupervised Mean Squared Error (uMSE), which is computed exclusively from noisy data. The uMSE is obtained by comparing the denoised image to an adjacent noisy frame and adding a correction term computed from two additional noisy frames (top row). If the signal content across the noisy frames is consistent and the noise is independent, the uMSE is an unbiased consistent estimator of the supervised Mean Squared Error between the denoised image and the underlying clean signal (bottom row). ***d) Denoising results:*** We compared the performance of a single-frame and multi-frame version of UDVD, against a traditional baseline based on Gaussian filtering (Gaussian) and an alternative unsupervised method known as Neighbor2Neighbor [4]. The metric was the unsupervised Peak Signal-to-Noise Ratio (uPSNR), which equals the logarithm of the uMSE. UDVD achieved a statistically significant superior performance for two datasets containing $CeO_2$ and platinum (Pt) nanoparticles. Supplement 1 provides additional details about the models and datasets.

For this investigation, we explore structural dynamics of Pt particles supported on $CeO_2$ in a CO environment at room temperature. CO interacts strongly with Pt surfaces, with a binding energy and migration energy of around 1.5 and 0.02 eV respectively [18, 19]. The CO surface coverage exceeds 50% even at the modest pressures of $10^{-4}$ to $10^{-2}$ Torr employed in the current experiment [20]. To investigate the dynamics, we recorded movies from a Pt/$CeO_2$ sample with an electron dose rate of 2000 $e^-$ $Å^{-2}$ $s^{-1}$ and a readout rate of 75 frames per second, corresponding to a single frame exposure time of 13 ms (individual frames had an electron dose of 26 $e^{-}$ $Å^{-2}$ and the dose per pixel was 0.2 $e^{-}$). Each movie is composed of approximately 1000 – 2000 frames (3500 x 3500 pixels in size) with a SNR (measured in the vacuum) of about 0.45, which obscures much of the surface structure in the raw data and makes it impossible to observe the underlying dynamics.

To process the low-SNR data, we leverage a deep-learning model trained and evaluated exclusively on the same real noisy data. The model is based on an Unsupervised Deep Video Denoiser (UDVD), recently developed by the authors [17, 21]. UDVD is trained to estimate each noisy pixel value using the surrounding spatio-temporal neighborhood, but without considering the noisy pixel itself (see ***Figure 1a***). This *blindspot* structure, which is enforced via a specialized architectural design, is critical, as it prevents the model from learning to trivially map the input to the output directly. Instead, the denoiser learns to estimate the underlying clean image structure without overfitting the noise. UDVD combines several UNet modules to process multiple frames at the same time, enabling it to exploit temporal patterns and multiscale structure (see ***Supplement 1*** for additional details). The results achieved by the denoiser are shown in ***Figure 1b***. After denoising, the atomic structure of the nanoparticles, including the surface, is clearly resolved, showcasing the advantage of unsupervised denoising for scientific discovery.

To evaluate the performance of UDVD, we apply an unsupervised evaluation metric recently developed by the authors: the unsupervised peak signal-to-noise ratio (uPSNR) [17]. This metric is computed using held-out adjacent noisy frames combined with a correction term (see ***Figure 1d***) that is guaranteed to yield an unbiased, consistent estimate of the true PSNR, under the assumption that the noise is independent across frames (this is approximately true, as shown in ***Supplement 1***). An additional qualitative evaluation of the denoised output was carried out by comparing a temporal average of the raw and denoised data. ***Figure 1c*** shows that there is reasonable agreement between the two temporal averages. Further details on training and evaluation of the denoiser output from generating nanoparticle surface structure are provided in ***Supplements 1 and 2***. Based on the vacuum region, the SNR in the UDVD output is approximately 26, which is improved by a factor of approximately 40 compared to the raw data. To achieve a similar improvement through counting statistics alone would require an increase in beam current or acquisition time by a factor of 1600. Increasing the beam current by such a large factor would destroy the material whereas increasing the acquisition time by this factor would destroy the time resolution. This demonstrates the power of the proposed denoising framework.

The denoiser reveals new dynamic behaviors on nanoparticle surfaces. ***Figure 2a-f*** shows a typical evolution of a 1.2 nm Pt nanoparticle surface supported on a (100) face of a $CeO_2$ during exposure to $10^{-4}$ Torr of CO at room temperature. The first image at t = 0s shows the particle in a (110) zone axis with crystallographic terminations corresponding to (111) surfaces. The particle undergoes rotation, and its shape evolves leading to the formation of a (100) facet. The presence of (111) and (100) crystallographic facets corresponds to the low energy Winterbottom shape for Pt nanoparticles [22]. The electron beam will always influence observations in the electron microscope. In this case, ***Supplement 3*** addresses this issue and compares energy transfers from the electron beam and thermal processes for Pt surface migration. The calculations show that thermally activated Pt jumps are $10^6$ times more likely than electron beam activated jumps, suggesting that the structural fluctuations are predominantly the result of thermal processes.

Interestingly, the high spatio-temporal resolution images show the presence of a diffuse contrast which appears to "float" above the crystallographic terminations. This component constantly changes in time and space and a layered chimney structure (labelled in ***Figure 2d***) is a pronounced example where, even though the nanoparticle is in a zone axis orientation with clearly resolved atomic columns, the chimney structure does not show atomic column contrast. This external surface layer is not an artifact of denoising and can also be seen (after suitable averaging) in the raw data (see ***Supplement 4***). Even low-energy (111) facets often have diffuse layers present a substantial fraction of the time. ***Figures 2g*** and ***h*** show an example from a different particle where the diffuse surface layer transforms to an ordered bulk terminated-like (111) Pt surface. This transformation implies that the diffuse layer, which we call an *adlayer*, is primarily associated with Pt atoms but the atoms are neither stationary nor in bulk terminated lattice sites. As ***Figure 2*** and associated images and movies in ***Supplement 4*** show, on small particles, the surface is constantly transforming between ordered crystallographic terminations and disordered adlayers. Sometimes the adlayer is associated with the nucleation or dissolution of a crystalline layer on the nanoparticle, whereas other times an existing crystalline layer transforms to an adlayer and then back to a crystalline layer as seen in ***Figure 2*** (and also in ***Figure 3***).

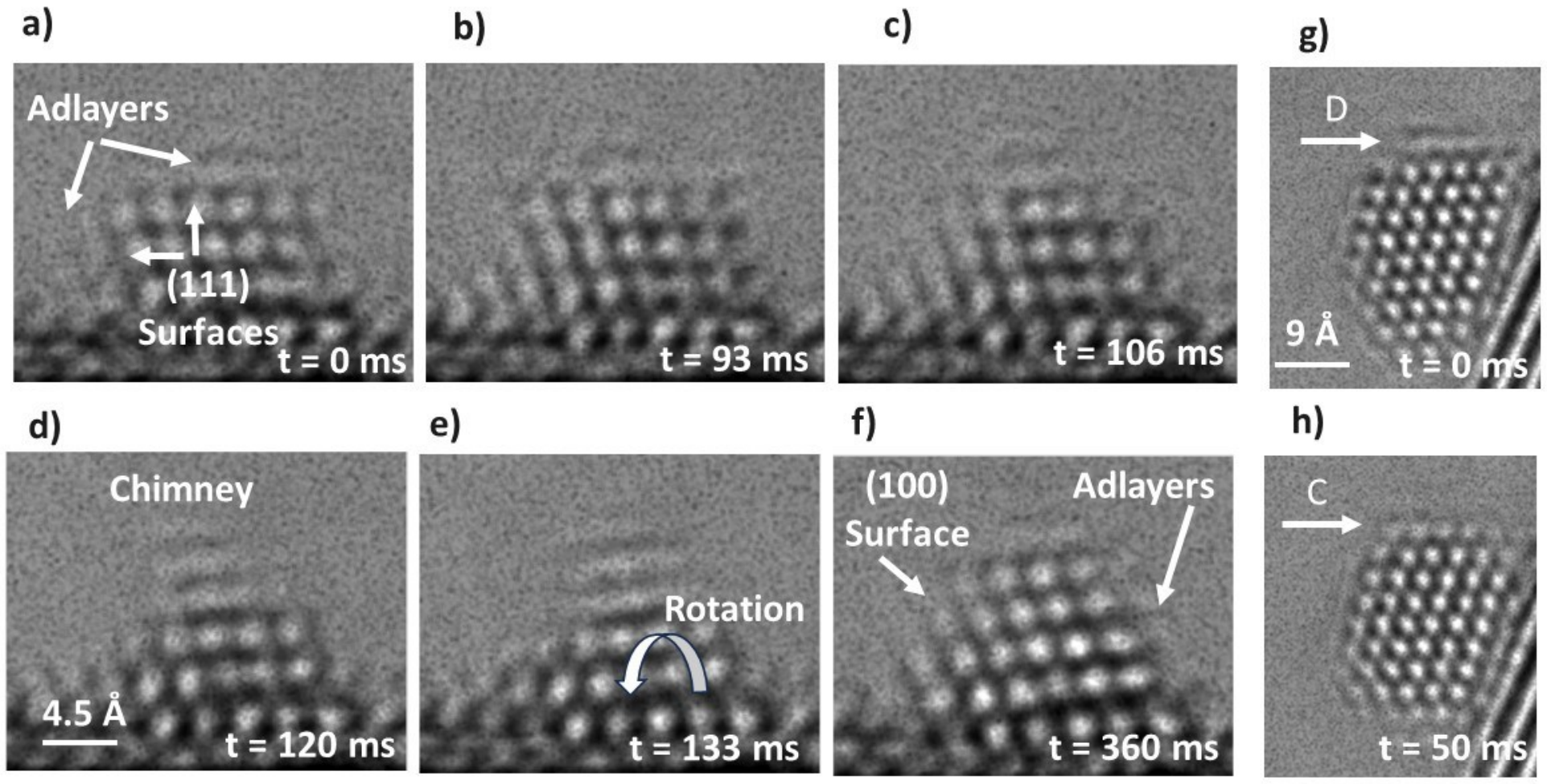

***Figure 2:*** **a) Surface Dynamics in Pt Particles a) – f):** The variation in the surface of a 1.2 nm Pt particle in a CO atmosphere over a time of 360 ms. The diffuse contrast at the surface of the nanoparticle are the dynamic adlayers of moving atoms. Rapid surface diffusion cause particle shape evolution such as the formation of metastable chimney structures and (100) facets. g)- h): 2 nm particle shows a disordered fluxional adlayer (D) transforming into a (111) crystallographic termination (C).

Adlayers composed of diffusing atoms have been reported in other areas of materials such as thin film growth and particle sintering via Ostwald ripening, but we are not aware that this phenomenon has been directly observed on nanoparticles. In Ostwald ripening of supported metal particles, adatoms diffuse on the metal surface and detach from the particles and migrate across the support to join other larger particles [23]. In the present case, the strong interaction with CO will disrupt metal-metal surface bonds increasing the likelihood that Pt atoms detach from lattice sites.

Moreover, most nanoparticles will not possess the correct number of atoms to form complete (111) and (100) layers to make the perfect Winterbottom shapes. This will increase the concentration of low coordination Pt atoms at step edges and corners sites making adatom detachment more facile. Once the atoms detach from crystal lattice sites, they are likely to be highly mobile. For example, the migration energy of Pt on Pt(111) surface is around 0.3 eV which would result in millions of jumps per second at room temperature [24] (see ***Supplement 3***).

The surface instabilities generate dynamic strain fields that penetrate below the surface and may trigger disruptions of subsurface layers. ***Figure 3*** captures the occurrence of a crystallographic shearing event taking place on a plane just below the surface leading to the formation of a stacking fault. In this case, a (111) Pt plane laterally slides, causing the ideal face centered cubic (fcc) stacking sequence (***Figure 3a***) to transform to a hexagonal close-packed (hcp) surface domain (see ***Figure 3e*** and ***Supplement 5***). Simultaneously, the particle undergoes a rigid body rotation of about 10$^{o}$ clockwise (making the (111) Pt plane parallel to the (111) surface of the $CeO_2$). The temporal evolution shows that the system passes through a transition state lasting about 13 ms (***Figure 3c***) during which the entire (111) plane shows streaked contrast characteristic of structural disorder or motion. Simultaneously, the adjacent surface layers on the left-hand side and upper left-hand side of the particle show adlayer contrast. This demonstrates that instabilities associated with surface adlayers can destabilize the crystal structure below the surface. The adlayers re-nucleate into ordered crystallographic terminations as the stacking fault is created (***Figure 3d*** and ***e***) and the structure stabilizes. The stabilization associated with stacking fault formation is short lived and the particle undergoes a reverse shear 600 ms later and the surface and subsurface layers then become very dynamic leading to the entire particle becoming fluxional, which manifests through rapid changes in atomic structure, particle shape and orientation (***Figure 3f-j*** and see movies in Supplement 4). The structure present in ***Figure 3j*** is not easy to determine. (The raw data shows very low contrast with a sparse rapidly changing phase contrast speckle suggesting rapid changes are occurring. The denoiser picks out the stronger features of the speckle pattern but these do not correspond to atomic columns).

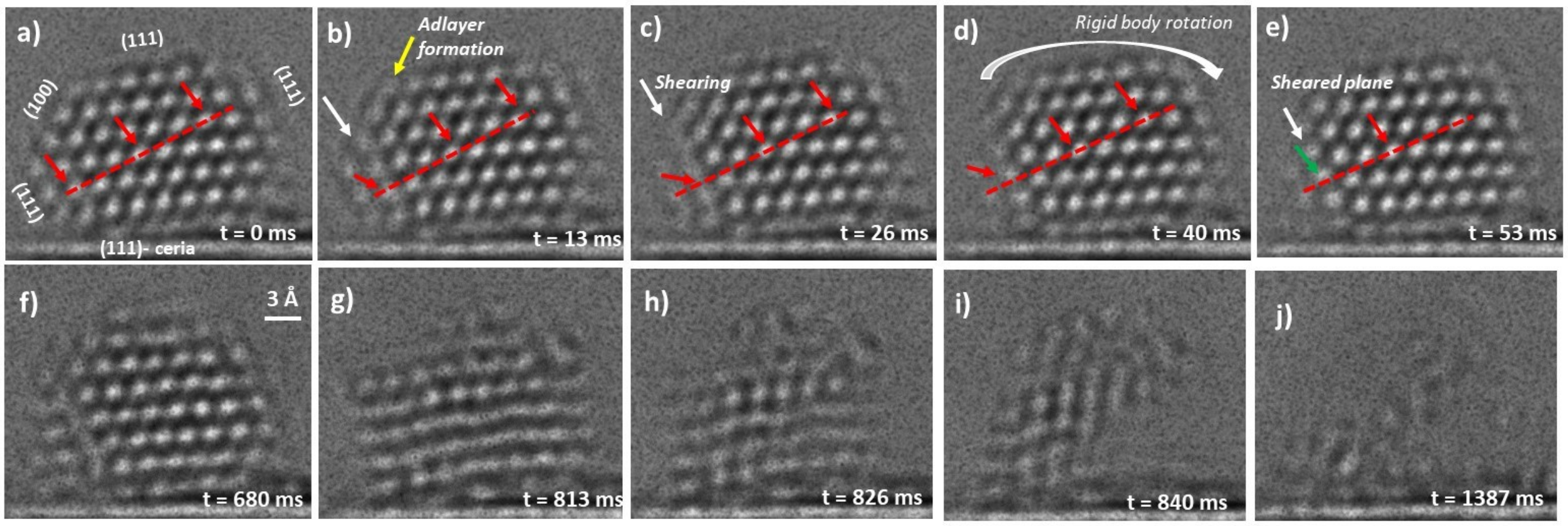

***Figure 3:*** **Dynamics at Subsurface Sites a) – e):** Sequence of images of Pt nanoparticle showing the formation of a subsurface stacking fault. **a)** shows the pre-sheared state of the nanoparticle. Red dotted line is perpendicular to a set of (111) planes with the bulk showing usual ABCA stacking and red arrows show the location of the A layers. **b)** the (111) plane (marked by white arrow) and (100) plane (marked by yellow arrow) shows streaking contrast demonstrating the onset of plane instability. **c)** the (111) and (100) dynamic adlayer formation indicating that pronounced atomic motion is occurring at rates beyond the frame exposure time of 13 ms. Rigid body rotation is also observed of the whole nanoparticle. **d)** and **e)** the (111) plane stabilizes in its new shifted position forming a stacking fault showing ABCB stacking (green arrow). The adlayers transform back to crystallographic terminations. **Entire Particle Fluxionality f) – j):** Surface interface instabilities drive structural dynamics and the phase contrast images become highly fluxional. The entire particle is destabilized resulting in rapid changes in crystal orientation and structure.

This whole particle fluxionality is more frequently observed in the smaller particles. One may hypothesize that the adlayer initiated disruptions below the surface are more likely when the surface-to-volume ratio is higher making small particles less stable. To explore this hypothesis in greater depth, we have developed an approach to quantify the order/disorder in images based on topological data analysis, specifically by means of summaries of persistent homology [25]. A brief description of the approach is given in ***Supplement 6***. Persistent homology essentially tracks the evolution in the degree of connectivity between dark (or light) pixels in a single image during intensity thresholding and, in a more ordered image, this connectivity is more persistent with thresholding. Specifically, this behavior is expressed via the so-called accumulated lifetime persistent survival (ALPS) statistic, which acts as a measure of structural order in the image. This summary is normalized for particle size in such a way that gives a value close to unity in the vacuum (corresponding to no order). An advantage of this approach is that it makes no assumptions about the form of the image (i.e. the presence of atomic columns, fringes etc.) so it can be applied to images from particles in any orientation. Applying this approach to a sequence of images from the same particle provides a quantitative way to compare how the order evolves with time (or to compare the degrees of order between nanoparticles).

***Figure 4a*** shows the ALPS statistic plotted as a function of time for the particle shown in ***Figure 3***. ALPS values of 1.3 or greater correspond to ordered structures whereas values of 1.1 or lower correspond to low degrees of order. The rapid small ALPS fluctuations of around 0.1 are not noise but are associated with constantly changing surface structures. ***Figure 4a*** provides a quantitative, high-level view of particle stability and explicitly shows the time that the system spends in metastable ordered states versus highly disordered states.

The ALPS statistics were employed to quantitatively compare the structural dynamics in nanoparticles of different size. It is applied to 23 movies (about 25,000 frames) from particles in the size range 0.7 to 4 nm (see ***Supplement 6*** for details). To simplify and facilitate the comparison between particles, the mean and standard deviations of each ALPS plot was determined and plotted as a function of particle size in ***Figure 4b*** and ***c***. The mean value of the ALPS statistic shows an approximately linear dependence with particles size (and surface-to-volume ratio) quantitatively confirming the hypothesis that instability is inversely proportional to size.

The standard deviation shows a poorly defined maximum in the size range 1.5 to 3 nm and suggests that there are three categories of structural dynamics for the Pt particles. The first category has the largest ALPS value (> 1.5) and relatively small standard deviations, corresponding to larger particles that remain well-ordered throughout the period of observation. Although their surfaces are dynamic, their ALPS statistic is dominated by the bulk (because of their small surface-to-volume ratio) and these particles remain relatively stable. The second category has the smallest ALPS values (<1.2) and small standard deviations. These are particles approximately 1.5 nm or smaller and they possess high degrees of disorder, and their low values of standard deviation shows that they are rarely in highly ordered states. From inspection of the denoised movies, the high degree of disorder is associated with high degrees of fluxionality. Since the particles have a high surface-to-volume ratio, fluxional surface adlayers drives fluxionality for the entire particle. The third category shows a wide range of standard deviations and a wide range of ALPS values. They are predominantly intermediate sized particles between 1.5 and 3 nm and these particles manifest very diverse behaviors and can either be extremely fluxional or relatively stable. The particle behavior depends on the degree of stability of their surfaces, and also on the stability of the interface with the support.

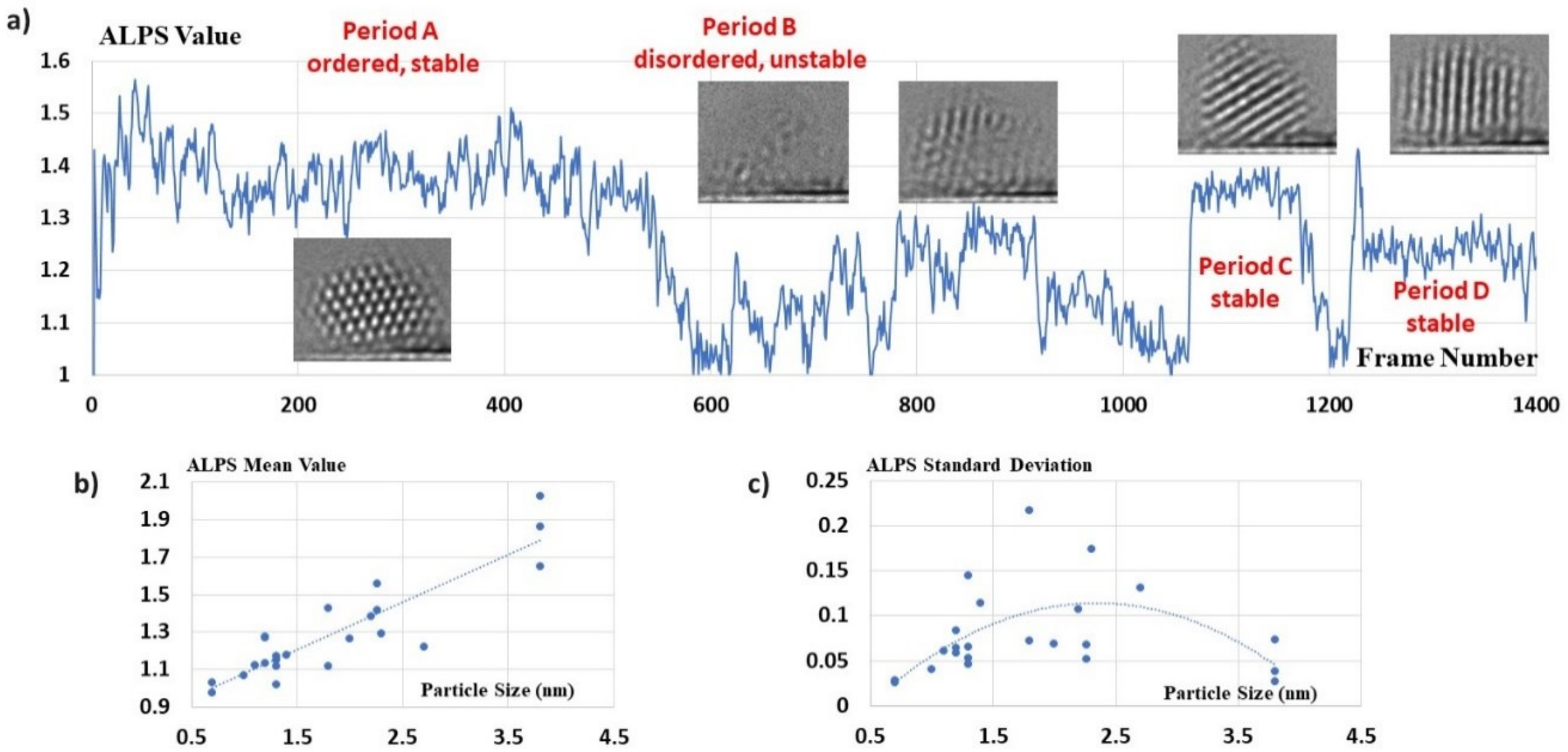

***Figure 4:*** **Quantifying Global Structural Dynamics in Pt Nanoparticles: a)** The order parameter (ALPS) versus frame number (frame time = 1/75 s ~ 13 ms) for the same particle showing the transition from ordered to disorders configuration (inserts are typical images for each of the four stable time periods A - D). b and c) The mean order parameter and standard deviation as a function of particle size for different nanoparticle measured with 13 ms time resolution over periods of 8 to 15 s.

The particle shown in ***Figures 3*** and ***4*** belongs to this third category which exhibits different degrees of order at different time periods. For period A (***Figure 4a***), the particle shows a well-defined orientation with the support. Activation of the reverse shear (***Figure 3f***) marks a period of more pronounced structural instability (period B), which manifest through large surface and interface changes. The availability of the denoised atomic resolution image for each ALPS point allows the structural origin for the stability/instability transformation to be explored. For example, the degree of instability is oscillatory throughout period B and inspection of the images shows that this is associated with a set of Pt (111) fringes (making an angle of 77° with the support) which repeatedly appear and disappear. The particle continuously attempts to establish a stable interfacial structure with the $CeO_2$ support but fails to achieve a stable configuration. The Pt particle surface is extremely fluxional throughout. The particle then enters more ordered period C, characterized by Pt(111) fringes making an angle of about 30° with respect to the support. Another shearing operation (at frame 1141) causes the particle to enter a brief period of instability before entering period D, an ordered stable period characterized by Pt(111) fringes making and angle of about 87° with respect to the support. The transitions from metastable to unstable configurations represent a rich and complex space but the topological analysis allows this complexity to be quantified in time and the denoised images permit the associated evolutionary structural pathways to be elucidated.

In summary, with the help of a newly developed unsupervised AI denoising algorithm and *in situ* electron microscopy, it is now possible to make atomic resolution observations of nanoparticle surfaces with time resolutions down to 10 ms and moderate electron dose. The structural dynamics of Pt nanoparticles in a CO atmosphere have been observed and characterized as a function of particle size. The nanoparticle surfaces continuously transition between relatively stable crystallographic terminations and more active adlayers composed of rapidly diffusing Pt atoms. The atoms of the adlayer temporarily "float" on top of the conventional crystallographic terminations and may nucleate to add a crystallographic monolayer to the surface or diffuse away. This process is continuous with the crystallographic terminations repeatedly stabilizing and destabilizing on timescales of less than 100 ms at room temperature. The surface structural dynamics and stress fields can penetrate below the surface leading to defect formation such as stacking faults. Many of the particles, especially the smaller ones, are observed to go through extended periods of extreme structural instability. Through the application of topological data analysis, we have been able to quantify and differentiate periods when the particle is in a well-ordered metastable state from the more fluxional disordered configurations. The high spatio-temporal information from the denoiser allows the short-lived atomic resolution elementary structural steps associated with nanoparticle transitions to be identified. The combination of AI-powered unsupervised denoising and *in situ* electron microscopy provides a new approach to investigate the field of atomic structural dynamics and stability. This will provide a new perspective for fundamental materials research by allowing functionalities to be correlated not only to static atomic structure, but also local structural dynamics.

**Acknowledgements**

We gratefully acknowledge financial support from the National Science Foundation. NSF OAC-1940263 and 2104105 supported PAC and PH, NSF CBET 1604971 supported JV, NSF DMR 184084 supported MT, NSF CHE 2109202 supported YW. NSF OAC-1940097 supported ML. NSF OAC-2103936 supported CFG. Support for DM and AT provided by NSF OAC-1940124 and DMS-2114143. The authors acknowledge ASU Research Computing and NYU HPC for providing high performance computing resources, and the electron microscopes in the Eyring Materials Center at Arizona State University. The direct electron detector was supported from NSF MRI 1920335.

# Supplement 1: Denoising Methodology

In this section we provide a detailed explanation of the denoising methodology used in this study, as well as additional results. [1]

## 1. Denoising Via Deep Learning

The goal of denoising is to address the presence of stochastic perturbations- typically known as "noise" that corrupt imaging data, obscuring key information. Over the last 15 years, deep learning has revolutionized computer vision and image processing, and denoising is no exception [lecun1995] [1]. Deep neural networks are the current state of the art for denoising on existing standard photographic-image benchmarks [2].

Deep neural networks compute complex, nonlinear functions of the input data by interleaving linear transformations with pointwise nonlinearities. In image processing, the neural networks are often convolutional, meaning that the linear transformations are implemented using spatially-invariant filters. A key consideration when processing large images, as is typical in electron microscopy, is the receptive field of the convolutional neural network (CNN), which is the region of the input image that is used to compute each output pixel. In this work, we leverage the U-Net architecture [3], originally designed for image segmentation, which achieves a large receptive field by incorporating several downsampling and upsampling layers. The name of the architecture is due to the skip connections that bypass each downsampling and upsampling level.

The standard approach for training denoising CNNs is minimizing an appropriate cost function via supervised learning [4]. In the case of denoising, the learning process leverages a training set of examples, consisting of pairs of noise and clean images. The cost function is the mean squared error (MSE) between the output of the network and the clean image corresponding to the noisy input. During training, the network parameters are modified iteratively to minimize the MSE, and therefore to approximate the clean images.

## 2. Unsupervised Denoising

The supervised-learning framework described in the previous section requires a database of ground-truth clean images, along with corresponding noisy measurements. In electron microscopy such databases are usually not available, especially when the goal is to uncover unobserved phenomena, as is the case in our study. In order to address this challenge, we have recently designed a CNN that can be trained exclusively on noisy data.

Our approach builds upon the blindspot method. In this approach, a CNN is trained to produce a denoised estimate that approximates the input noisy data. This sounds very naïve: the network can simply implement the identity data and output the noisy pixels! The key insight is that this can be avoided by *blinding* the CNN, so that it estimates each denoised pixel from its surrounding spatial neighborhood, but *without including the noisy pixel itself*. Assuming the noise is pixel-wise independence, then the network cannot replicate the noisy component in the data (as it does not observe it due to the blindspot mechanism), and is consequently forced to approximate only the underlying clean image. Originally, the blindspot framework was implemented as the Noise2Void approach, where pixels are masked by replacing them with random values [5] (see also [6] for a similar method known as Noise2Self). Subsequently, a specialized architecture was designed to explicitly create a blindspot in the receptive field [7].

Our denoising CNN, which we call Unsupervised Deep Video Denoiser (UDVD), is a modification of the blindspot architecture that is able to process multiple frames. UDVD maps five contiguous noisy frames to a denoised estimate of the middle frame. The architecture is designed so that each output pixel is estimated from a spatiotemporal neighborhood that excludes the pixel itself. This is achieved by rotating the input frames by multiples of 90 degrees and processing them through four separate branches consisting of asymmetric convolutional filters that are vertically causal. As a result, the branches produce a denoised pixel that only depends on the pixels above (0° rotation), to the left (90°), below (180°) or to the right (270°).

Each branch in UDVD consists of two stages, inspired by previously proposed networks for supervised video denoising [8, 9]. The first stage, consisting of three UNets with shared parameters, maps each group of three contiguous frames (i.e. (t-2, t-1, t), (t-1, t, t+1) and (t, t+1, t+2)) to a different feature map. These features are then mapped to a single feature map using another UNet. The outputs of the four branches in UDVD derotated, so that they align, and combined using a three-layered cascade of 1x1 convolutions and nonlinearities to produce the final denoised estimate output.

The network parameters are trained by minimizing the mean squared error between the denoised output and the noisy input images. Early stopping is performed based on a validation set consisting of additional held-out noisy images. More details, and a detailed analysis of the performance of UDVD on different types of video data, are reported in our conference publication [10].

## 3. Unsupervised Evaluation

In the deep-learning literature, the evaluation of unsupervised denoising methods has relied on images and videos corrupted with synthetic noise [5-10], which are not available in many real-world scenarios, including this study. In order to address this, we have recently proposed a framework for unsupervised evaluation of denoisers, which relies exclusively on noisy data. The key idea is to compare the denoised signal to a noisy reference, which corresponds to (approximately) the same clean signal corrupted by independent noise.

In order to explain our approach, let us consider an image corrupted by additive noise, so that the noisy measurements equal $y := x + z$, where $x$ is the clean image and $z$ denotes zero-mean independent noise (here $y$, $x$ and $z$ all have the same dimension, equal to the number of pixels). Assume that we have access to a noisy reference $a := x + w$, corresponding to the same underlying clean image $x$, but corrupted with a different noise realization $w$ independent from $z$ (below, we explain how to obtain such references in practice). Let $f(y)$ denote a denoised estimate, obtained by processing $y$. The mean squared error (MSE) between the denoised estimate and the reference is approximately equal to true MSE between the clean image and the denoised estimate, summed with the variance of the noise:

$$\frac{1}{n}\sum_{i=1}^{n}(a_i - f(y)_i)^2 = \frac{1}{n}\sum_{i=1}^{n}(x_i + w_i - f(y)_i)^2$$

$$\approx \frac{1}{n}\sum_{i=1}^{n}(x_i - f(y)_i)^2 + \frac{1}{n}\sum_{i=1}^{n} {w_i}^2 \ (1)$$

Here n denotes the number of pixels. The terms $2w_i(x_i - f(y)_i)$ approximately cancel out when summed, as long as the new noisy realization $w_i$ and $f(y)_i$ are independent (which is the case if $w_i$ and $f(y)_i$ are independent).

Approximations to the equation above are used by different unsupervised approaches to train neural networks for denoising [11]. The noise term $\frac{1}{n}\sum_{i=1}^{n} {w_i}^2$ is not an obstacle for training denoisers, if it is independent from the input y. However, it is definitely problematic for evaluating denoisers, as the additional term is different for different images and datasets, so it cannot be used for quantitative comparisons. In order to address this limitation, we modify the cost function to cancel out the noise term. To this end, we utilize two additional references b:=x + v and c:=x+u, which are noisy images corresponding to the same clean image x, but corrupted with different, independent noise realizations v and u (just like a). Subtracting these references and dividing by two yields an estimate of the noise term,

$$\frac{1}{n}\sum_{i=1}^{n}\frac{(b_i - c_i)^2}{2} = \frac{1}{n}\sum_{i=1}^{n}\frac{(v_i - u_i)^2}{2} \approx \frac{1}{n}\sum_{i=1}^{n}\frac{{v_i}^2}{2} + \frac{1}{n}\sum_{i=1}^{n}\frac{{u_i}^2}{2} \approx \frac{1}{n}\sum_{i=1}^{n} {w_i}^2$$

under the assumption that the noise is pixel-wise independent and all the noisy perturbations have the same distribution. We subtract this noise estimate from the quantity in equation (1) to estimate the MSE. This yields our proposed unsupervised metric, which we call unsupervised mean squared error (uMSE).

Given a noisy input image y and three noisy references a, b, c, the unsupervised mean squared error of a denoiser f is

$$uMSE := \frac{1}{n}\sum_{i=1}^{n}(a_i - f(y)_i)^2 - \frac{1}{n}\sum_{i=1}^{n}\frac{(b_i - c_i)^2}{2}.$$

In our conference publication [12], we establish that the uMSE is a consistent estimator of the MSE as long as (1) the noisy input and the noisy references are independent, (2) their means equal the corresponding entries of the ground-truth clean signal, and (3) their higher-order moments are bounded. These conditions are satisfied by Poisson shot noise, which is often the dominating source of noise in electron microscopy data acquired with direct electron detectors (the present case).

In image processing, it is common to measure denoising quality using a logarithmic function of the MSE called the peak signal-to-noise ratio (PSNR), defined on a decibel scale

$$PSNR := 10\, log \frac{M^2}{MSE},$$

where M is a fixed constant representing the maximum possible value in the images. The uMSE can be naturally extended to yield an unsupervised PSNR (uPSNR):

$$uPSNR := 10\, log \frac{M^2}{uMSE}.$$

The uPSNR is a consistent estimator of the PSNR, under the same conditions that guarantee consistency of the uMSE.

## 4. Computing Noisy References From Microscopy Data

The proposed unsupervised metrics described in the previous section require three noisy references, which should correspond to the same clean image contaminated with independent noise. A possible strategy to obtain such references from noisy video data is spatial subsampling. In spatial subsampling each frame is partitioned into 2x2 blocks. The pixels in each block are then randomly assigned to each reference. Under the assumption that the noise is independent among adjacent pixels, and the underlying clean image is smooth with respect to the pixel resolution, this yields four noisy references approximately satisfying the assumption of the unsupervised metrics. This is the approach taken in our conference publication [12].

Here, we propose a different subsampling scheme to obtain our noisy references: temporal subsampling. For each frame, we use the three nearest frames as noisy references. Assuming the noise is independent from frame-to-frame and the underlying clean-signal dynamics are sufficiently smooth with respect to the frame rate, this yields four noisy references approximately satisfying the assumption of the unsupervised metrics.

The reason that we chose temporal subsampling over spatial subsampling for our dataset of interest is that the noise correlation between adjacent frames is weaker between adjacent frames, as opposed to between spatially-adjacent pixels within the same frame. ***Figure S1A*** shows the spatial and temporal correlation for two datasets containing $CeO_2$ and platinum (Pt) nanoparticles (see Section 5). The correlation is computed using pixels corresponding to vacuum, where the measured signal is exclusively due to noise. We observe that the inter-frame correlation for both datasets is around $10^{-3}$ or smaller. In contrast the correlation between spatially-adjacent pixels is one order of magnitude higher. Consequently, temporal subsampling is more consistent with the assumptions of the unsupervised metrics.

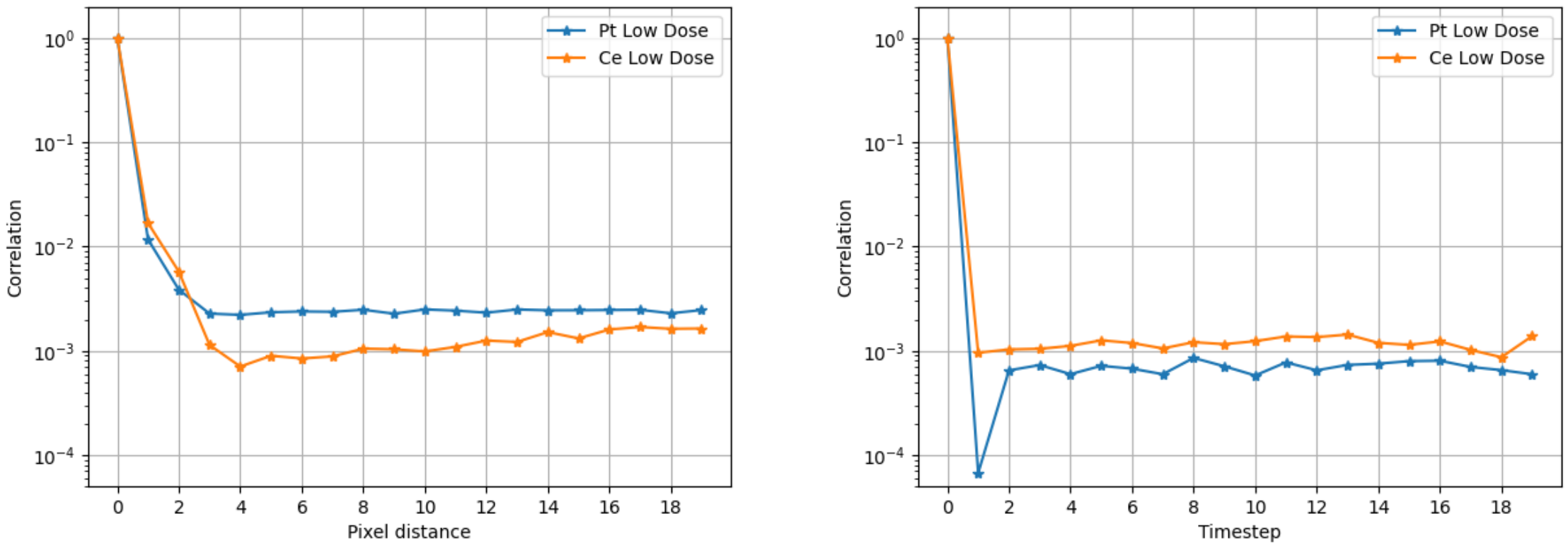

***Figure S1A***: Empirical correlation between noisy pixels extracted from the vacuum regions of two datasets containing $CeO_2$ and platinum (Pt) nanoparticles (see Section 5). The correlation between adjacent frames (right graph, timestep = 1) is $10^{-3}$ or smaller. The correlation between spatially-adjacent pixels is one order of magnitude higher (left graph, pixel distance = 1).

## 5. Experiments

In order to evaluate the proposed methodology, we have utilized several datasets (movies) that focus on different materials systems. The first dataset (A) is of direct relevance to the current manuscript and comes from Pt nanoparticles supported on a $CeO_2$ support exposed to a CO atmosphere. All of the images shown in the current manuscript are drawn from this dataset. We have also included evaluation results from a second related but different dataset (B). This data comes from $CeO_2$ nanoparticles tilted into zone axis orientation so that the atomic resolution images reveal anion (oxygen) and cation (ceria) columns. The focus of this project is to understand the dynamics of reducible oxide surfaces. The characteristics of this second dataset are quite different from the Pt dataset which allows us to evaluation the methodology from contrasting materials. Specific details of the data acquisition for each data set are given below:

A) Platinum: The platinum on ceria movie used for training and validation was acquired with an electron dose rate of 2000 $e^- Å^{-2} s^{-1}$ and a readout rate of 75 frames per second, corresponding to a single frame exposure time of 13 ms (individual frames had an electron dose of 26 $e^- Å^{-2}$ and the dose per pixel was 0.2 $e^-$). Each movie is composed of approximately 1000 – 2000 frames (3500 x 3500 pixels in size) with a SNR (measured in the vacuum) of about 0.45. This is the dataset that was employed for the main manuscript analysis of Pt nanoparticle fluxionality.

B) $CeO_2$: The pure ceria movie used for training and validation was acquired with an electron dose rate of 3900 $e^- Å^{-2} s^{-1}$ and a readout rate of 7.5 frames per second, corresponding to a single frame exposure time of 130 ms. The movie is composed of approximately 800 frames (3500 x 3500 pixels in size) with a SNR (measured in the vacuum) of about 1.26. No experimental data from this second dataset is included in the main manuscript.

We compared the performance of our proposed methodology (UDVD, described in Section 2) to several baselines on the two datasets: (1) a single-frame version of UDVD, which uses one noisy frame instead of five, (2) an alternative unsupervised method known as Neighbor2Neighbor [13],

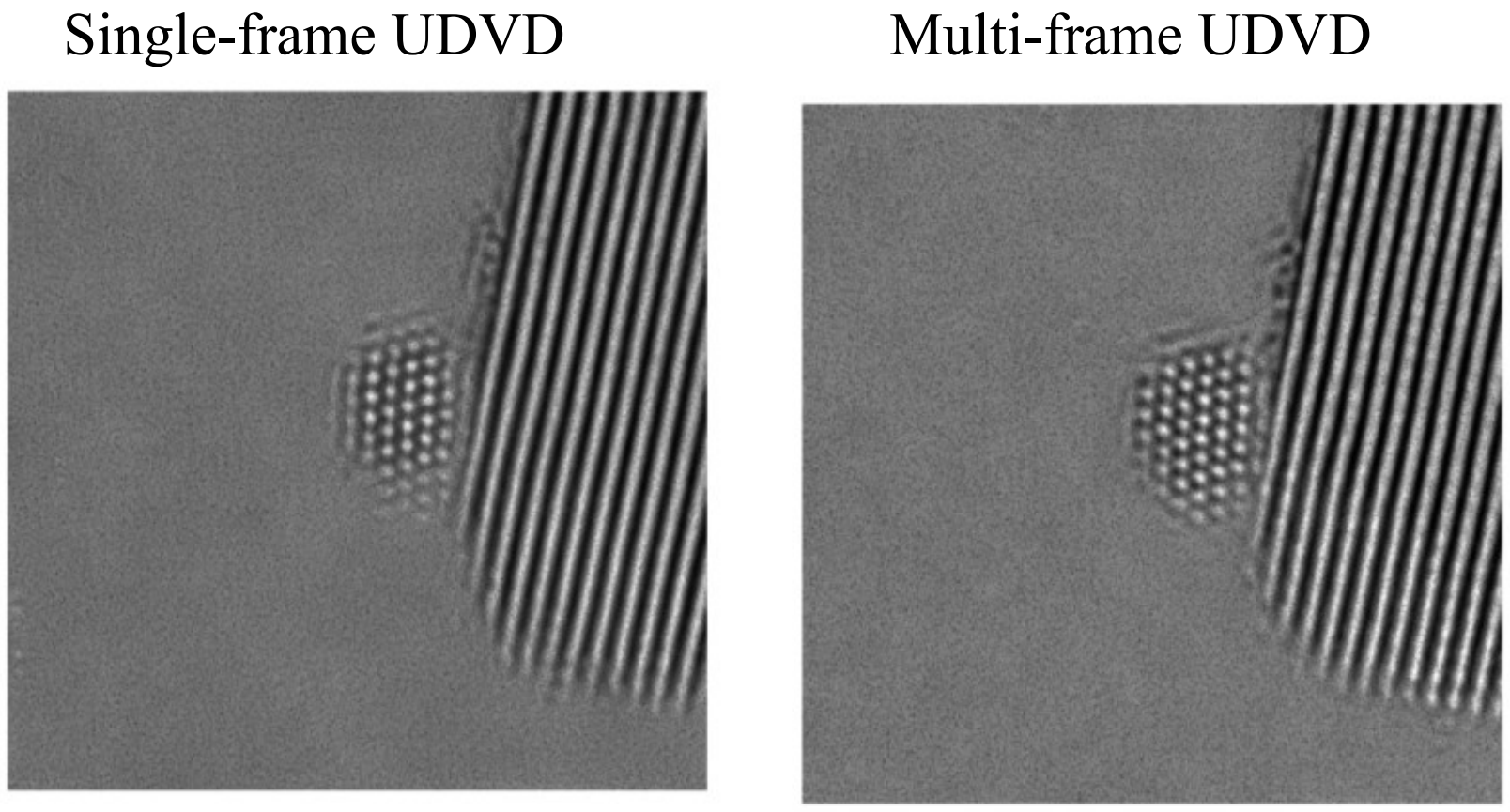

Figure 2: Comparison between the multi-frame and single-frame versions of the proposed unsupervised deep video denoiser (UDVD) applied to the Pt nanoparticle data. Superficially, the single frame and multiframe look similar. But closer inspection reveals that the single frame shows a higher number of phantoms in the vacuum (and surface) and erratic changes in the bulk structure between adjacent frames suggesting a greater percentage of errors in up sampling.

(3) a traditional Gaussian filtering denoising method (sigma = 2 pixels applied) available in the ImageJ software [14], which was optimized by a domain expert. The deep-learning methods were trained on the first 600 frames from the noisy datasets for 500 epochs with a batch size of 2 using the Adam optimizer [15]. A random sample of size 512 × 512 from each frame was used for training. All methods were evaluated on the last 150 test held-out frames using the unsupervised Peak Signal-to-Noise Ratio (uPSNR, described in Section 3) based on temporal subsampling (see Section 4). 90% confidence intervals for the uPSNR metric were computed via bootstrapping (see Appendix A in [uMSE]). Multi-frame UDVD achieved a superior performance over the baselines for both datasets. The differences are statistically significant, except for the difference between single-frame and multi-frame UDVD for the platinum dataset. However, visual inspection indicates that the multi-frame output is indeed qualitatively superior, as illustrated by ***Figure S1B***.

## Supplement 2: Strategies for Denoising Movies of Pt Nanoparticles

The ability of the denoiser to return a movie that is close to the "ground truth" will depend on its ability to learn from the information that is present in the raw data. If too little data is supplied the output from the network will show artifacts. We have explored the effect of changing parameters such as signal-to-noise, patch size, frame size, number of frames, numerical precision, image region, and drift correction. The effect of these parameters on the output from the network will be presented and discussed in detail in a future publication. Here we show that with optimization of these parameters, we can generate and output relatively free from artifacts with reasonable computation cost.

If we consider just the Pt particle, the ground truth would require knowledge of the structure and orientation of the Pt particles at each point in time. Clearly this is an unknown. However, the field of view contains vacuum as well as the Pt nanoparticle. In a phase contrast bright-field image, the vacuum shows up with uniform, constant intensity at all locations a nanometer or more away from the particle surface (near the particle surface it is possible to observe Fresnel fringe depending on the defocus value). Thus we effectively can use the vacuum region of the image as a region of ground truth. A necessary condition for the denoiser to perform well is that the vacuum should appear with uniform contrast in the output. This criterion is extremely useful for constraining the parameters for denoising movies. If the number of frames or frames size is insufficient, we observe artifacts in the vacuum including phantom atoms and fringes. It is helpful to measure the uniformity of the vacuum both in real space and in the Fourier transform on the image.

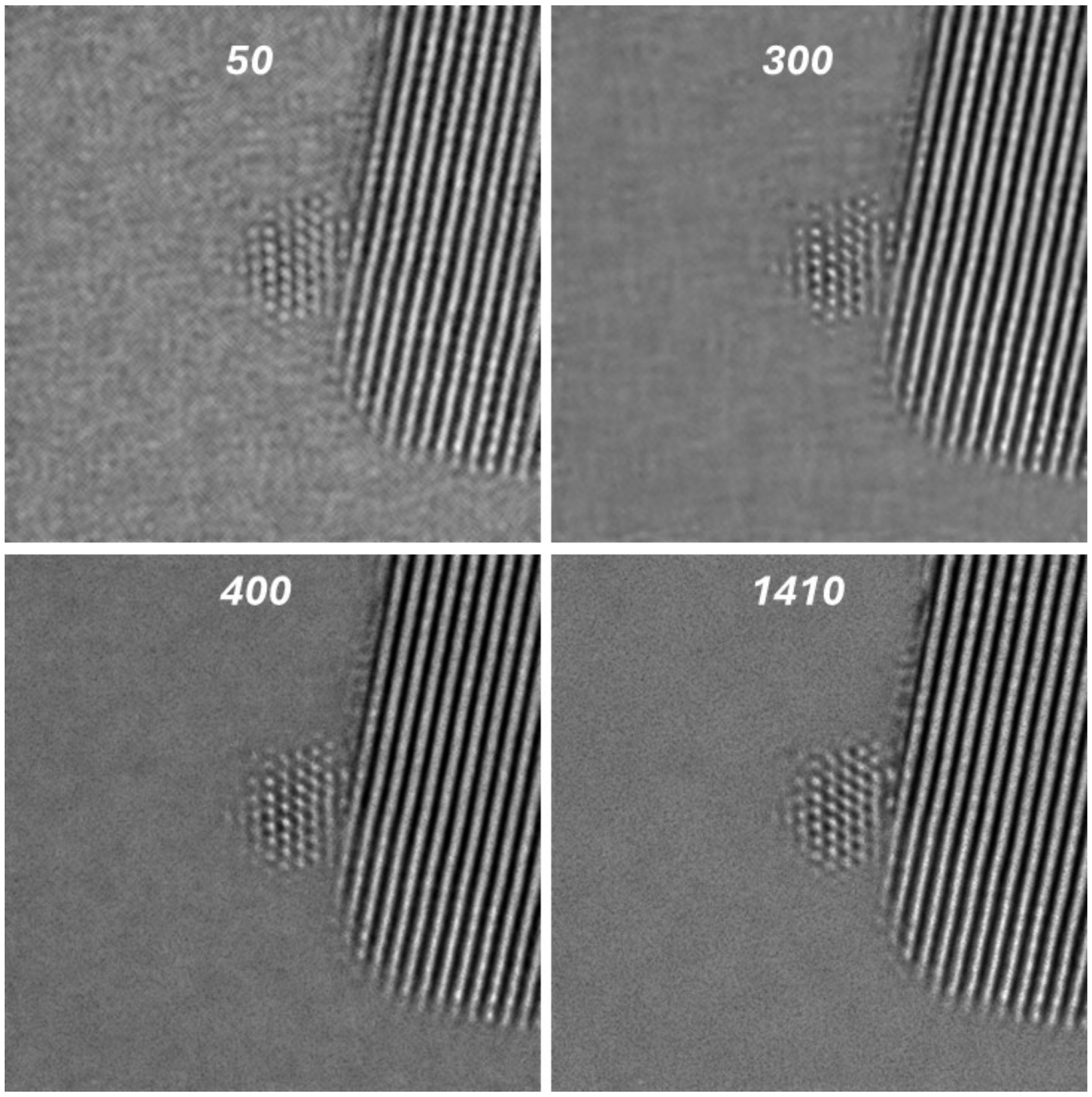

***Figure S2A:*** Denoised images of the same frame after training the denoiser with inputs of 50, 300, 400 and 1410 frames.

**Figure S2A** shows examples of the output from the denoiser for inputs of 50, 300, 400 and 1410 frames (keeping the frame size constant at 1024 x 1024). The output from a 50-frame input segment (computation time 0.07 hour) is dominated by severe artifacts in the vacuum effectively obstructing the view of the nanoparticle surface. The Fourier transform shows that these artifacts have spacings that are characteristic of the $CeO_2$ and Pt lattices. The network has learned the characteristic motifs from the Pt and $CeO_2$ structures and is imposing them on the vacuum noise. These vacuum artifacts are reduced but still prominent when the number of frames is increased to 300. By 400 frames, there is a significant reduction in the structure in the vacuum and by 800 frames is not detectable. Increasing to 1410 frames (the entire length of this particular movie), there is very little difference in the appearance of the real space vacuum but there is a continued improvement in the Fourier transform. One can conclude that for the specific conditions employed in the current experiment, movies recorded with at least 800 frames is adequate to reduce vacuum artifacts to an acceptable level. (It is worth commenting that the training is conducted with only 2/3 of the frames, the other 1/3 are reserved for validation. So when the network is provided with input of say 400 frames, only 266 frames are actually used for training with the remaining 134 being used for validation).

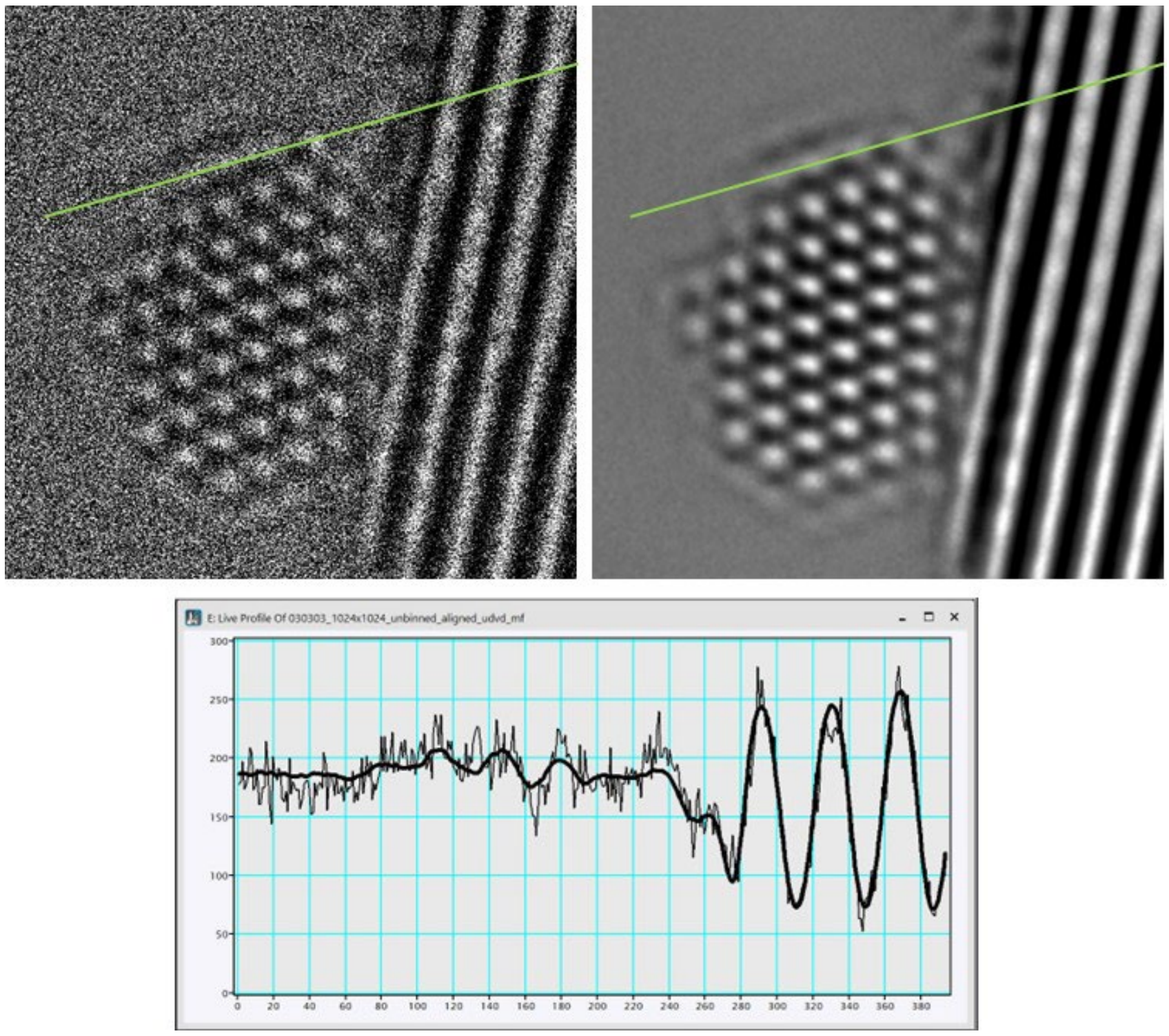

***Figure S2B:*** Raw data (left) and denoised data (right - after training with input of 1410 frames) of 30 frame sum. Linescan (green linefrom denoised (smooth line) and raw data (noisy line) showing similar intensity variations. The time-averaged surface structures look very similar in both cases.

We have no direct way of determining if the surface structure revealed by the denoiser is 100 % correct since the SNR of the input raw data is very poor. However, if the denoiser is performing well, then summing and/or binning output frames should give an image which is very similar to the image obtained by applying the same summing and/or binning operation to the raw data. This is illustrated in **Figure S2B** where 30 frames have been summed in the raw data to improve the SNR from and the same frames are summed in the output from the denoiser. Binning the raw data by a factor of 30 increases the SNR from 0.45 to 2.5 and time-averaged surface structure becomes visible. Comparing the surfaces in both images shows that they are very similar providing a high degree of confidence that the surface image obtained from the network is reasonably close to the ground truth. The conclusion is also supported in Supplement 4, where the image contrast from specific types of defects are shown to be similar in both raw data and denoised data.

# Supplement 3: Comparing Migration Rates from Electron Beam and Thermal Excitation

In electron microscopy, the fast electron beam can transfer energy to a Pt surface atom and potentially drive surface migration. The electron beam can transfer energy to the atom through either ionization processes (radiolytic displacement) or direct collisions with the atomic nucleus (knock-on damage) [1-3]. The rate of Pt surface jumps caused by the electron beam can be estimated from electron scattering cross sections. A necessary condition for electron beam induced surface migration is that the energy transferred from the electron beam to the Pt atom must be greater than or equal to the activation energy for surface migration, $E_a$. By calculating cross sections arising from knock-on and radiolytic displacements, we can estimate the oxygen displacement rate due to beam damage as a function of $E_a$.

## Knock-on induced Surface Migration

The following derivation is based on that of Egerton, Wang, and Crozier for electron beam damage from an intense probe [4]. Knock-on displacement damage may be considered to be an elastic scattering process. During elastic scattering by an atom, an electron that undergoes an angular deflection θ transfers to the nucleus an amount of energy equal to:

$$E = E_{max} sin^2(\theta/2)$$

where $E_{max}$ is the maximum possible energy transfer, corresponding to θ = π rad. Relativistic kinematics gives:

$$E_{max} = 2E_0(E_0 + 2m_0c^2)/Mc^2$$

where M is the mass of the scattering atom, assumed initially at rest, and $E_0$ is the kinetic energy of the incident electron (rest mass $m_0$); c is the speed of light in vacuum and $m_0c^2$ = 511 keV is the electron rest energy. For Pt, M = 195g /6.022x$10^{23}$ = 3.238 x $10^{-22}$ g = 3.238x $10^{-25}$ kg / 1.6 x $10^{-19}$ C = 2.02 x $10^{-6}$

For Pt: $$E_{max} = \frac{2(300kV)[300kV+2*511kV]}{[(2.02\ x\ 10^{-6})(2.998\ x\ 10^{8})^2]} = \frac{[7.932\ x\ 10^{11}]}{[1.815\ x\ 10^{11}]} = 4.37\ eV$$

$E_{min} = E_a$ (Pt migration energy) is a variable in our derivation to produce an equation where the displacement rate depends on activation energy.

Neglecting screening of nuclear field, which is a good approximation for large scattering angles, the differential cross section for such Rutherford-type scattering is:

$$\frac{d\sigma}{d\theta} = \left[\frac{e^2Z}{(8\pi\varepsilon_0E_0)}\right]^2 \left[\frac{E_0 + m_0c^2}{E_0 + 2m_0c^2}\right]^2 * \left[\frac{2\pi sin\theta}{\sin^4\left(\frac{\theta}{2}\right)}\right]$$

This expression can be integrated over the scattering angle, from θ = π to a minimum value given by $\sin^2(\theta/2) = E_{min}/E_{max}$, to give a cross section for energy transfer in the range $E_{min}$ to $E_{max}$:

$$\sigma = (2.45\ x\ 10^{-29} m^2) Z^2 \left[ \frac{1 - \frac{v^2}{c^2}}{\left(\frac{v^2}{c^2}\right)^2} \right] * \left[ \left(\frac{E_{max}}{E_{min}}\right) - 1 \right]$$

For Pt: $\sigma = (2.45\ x\ 10^{-29} m^2)(78)^2 \left[ \frac{\left(1 - \frac{(2.33\ x\ 10^8)^2}{(2.998\ x\ 10^8)^2}\right)}{\left(\frac{(2.33\ x\ 10^8)^2}{(2.998\ x\ 10^8)^2}\right)^2} \right] * \left[ \left(\frac{4.37 eV}{E_a}\right) - 1 \right]$

This yields and expression for the diffusion cross section due to elastic scattering as:

$$\sigma = (1.617\ x\ 10^{-25}) \left[ \left(\frac{4.37 eV}{E_a}\right) - 1 \right]$$

For a flux of $D$ electrons per unit area, and damage cross section $\sigma$, the number of migration events per unit surface area, $n$, is given by $n = D\ \sigma\ n_{at}$ where $n_{at}$ is the number of surface atoms per unit area. A more general quantity that can be used for critical flux calculations is the number (or fraction) of displacement events per target atom,[5] $x = n/n_{at} = D\ \sigma$.

In our experiments, a flux of 2000 $e^- Å^{-2} s^{-1}$ was used, and assuming Pt atom is about 2.5 A in diameter which gives the following formula for the number of migration events per second.

**Pt migration rate:** $\frac{n}{n_{at}} = D\sigma = \left(5x\ 2000 \frac{e^-}{Å^2 s}\right) * (1.617\ x\ 10^{-25}) \left[ \left(\frac{4.37 eV}{E_a}\right) - 1 \right]$

This is shown by the blue line in ***Figure S3A.***

## Radiolytic Induced Surface Migration

The following derivation follows that of Hobbs [2]. Radiolytic displacement may be considered to be an inelastic scattering process. The inelastic displacement cross section is:

$$\sigma = 7x10^6\ \xi \left(\frac{Z}{E_a}\right) * (10^{-28})$$

$\xi$ is the efficiency factor and is typically an empirically derived value. For silica, the value is $10^{-4}$. but for metals it is many orders of magnitude lower because of screening by the conduction band electrons. For this reason, we neglect surface migration induced by radiolysis.

## Thermally Induced Surface Migration

The surface atoms will also migrate due to thermal fluctuations. The number of jumps per second of a surface atom , $n_{th}$, can be roughly estimated using the Arrhenius equation as:

$$n_{th} = A\exp(-E_a/kT)$$

where A is the attempt frequency, T is temperature and k is Boltzmann's constant. The attempt frequency can be approximated as the phonon frequency of 1 x $10^{12}$ Hz. The value of $n_{th}$ is also plotted as a function of $E_a$ on ***Figure S3A.***

Comparing the migration rate due to the electron beam with that due to thermal fluctuation, ***Figure S3A*** show that for migration energies of less than 0.7 eV, thermal effects dominated whereas electron beam effects dominate for migration energies above this value. To determine the importance of electron beam effects on the surface dynamics, it is necessary to know the activation energies for surface process. The migration energy for Pt migration on (111) Pt surface is about 0.25 eV and such a process if more likely to occur due to thermal process (which are 7 orders of magnitude more likely than the electron beam effect). Detachment of atoms from kink sites on steps is one possible mechanism for creating adatoms. The are many way in which this can happen with a wide range of activation energies. However, especially in CO, calculations suggest that many of the processes have activation energies of 0.5 eV or lower which will be dominated by thermal effects [5, 6]. This appears to be true not only for Pt but also other metal such as Cu [7, 8]. Detachment of a Pt atom from a complete step is close to 1 eV giving an activation rate of about $10^{-5}$/s at room temperature. In this case the electron activation rate is much higher but is still only about 0.5/s. Thus in comparison to all the over adatom activation processes, this will make a negligible contribution to the adatom population.

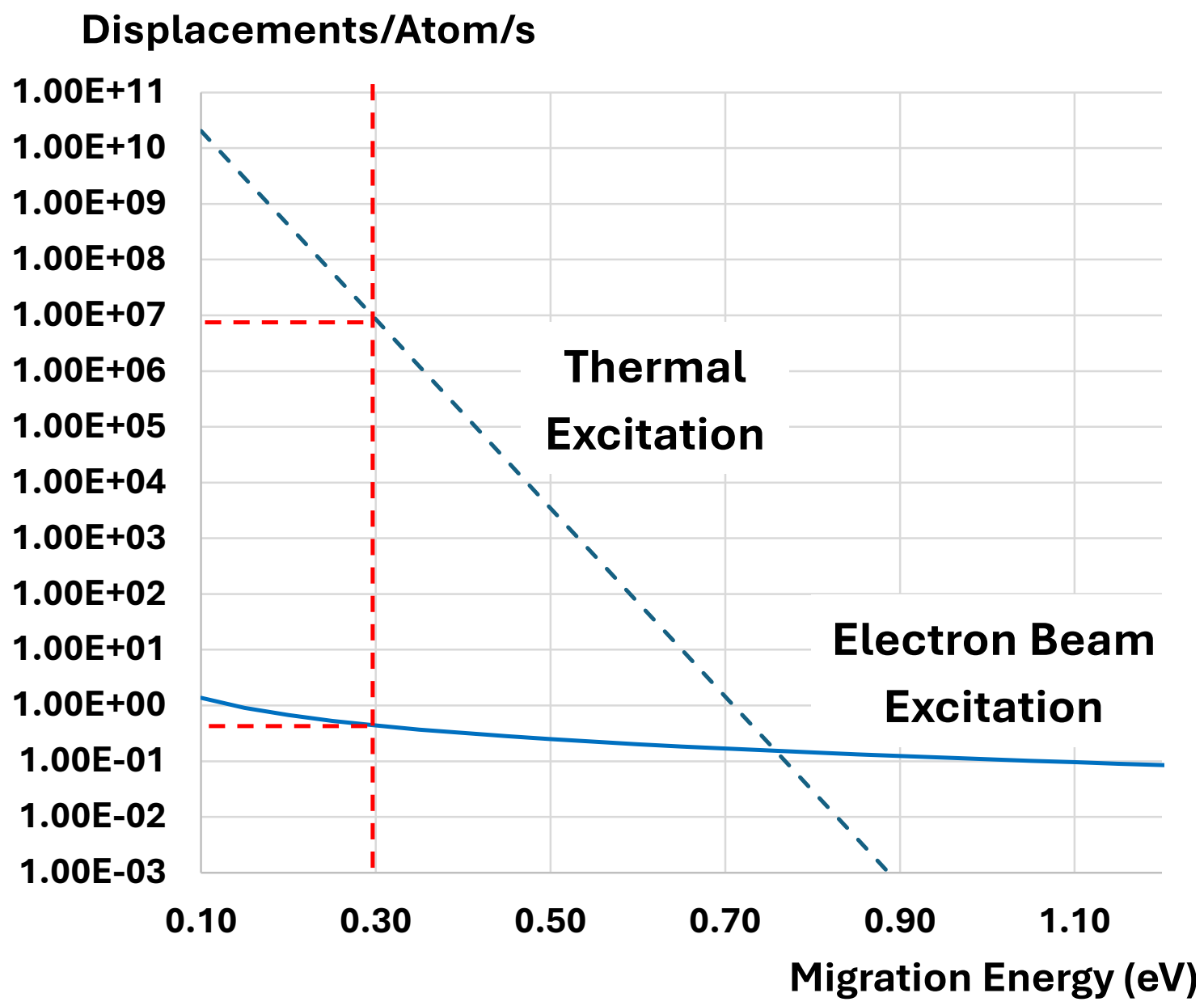

***Figure S3A:*** The number of migration jumps per second as a function of activation energy due to electron beam and thermal excitation. The migration energy of Pt on (111) surface is 0.3 eV and jump rate marked with dotted red line.

So in summary, by combining activation energy for Pt migration and atom detachment in a CO atmosphere from calculations with electron knock-on cross sections and thermal excitation probabilities, we conclude that electron beam effects will be many orders of magnitude less important than thermal process. Thus, the observed surface dynamics is dominated by thermal excitation processes.

## Low Dose Observations

To further validate that the observation are not primarily due to electron beam effects we repeated experiments at an incident dose of 200 $e^{-}Å^{-2}s^{-1}$, i.e. a factor of 10 times lower dose. For this condition, the calculations above suggest that the number of Pt surface atoms displacements by knock-on damage will be around 1 displacement every 20s. The thermal displacement should remain the same at around 1 x $10^7$ displacements per second. For this electron dose rate, the average image intensity was only electrons per pixel was ~0.045 e/pixel (at 75 frames per second) giving a vacuum SNR of 0.2. The current set of parameters employed in the UDVD algorithm fails at such low doses, so it is not possible to investigate structural dynamics on the nanoparticle surface at time resolution of 13 ms. However, by frame averaging the raw data, we can see changes in the image contrast as a function of time. ***Figure S3B*** show three images from a 2 nm Pt nanoparticle separated in time by about 15 s. 35 frames have been summed to give an exposure time of 0.5 s which is adequate to see atomic resolution contrast. The contrast pattern is different in the three images showing that rotation and/or structural change is taking place.

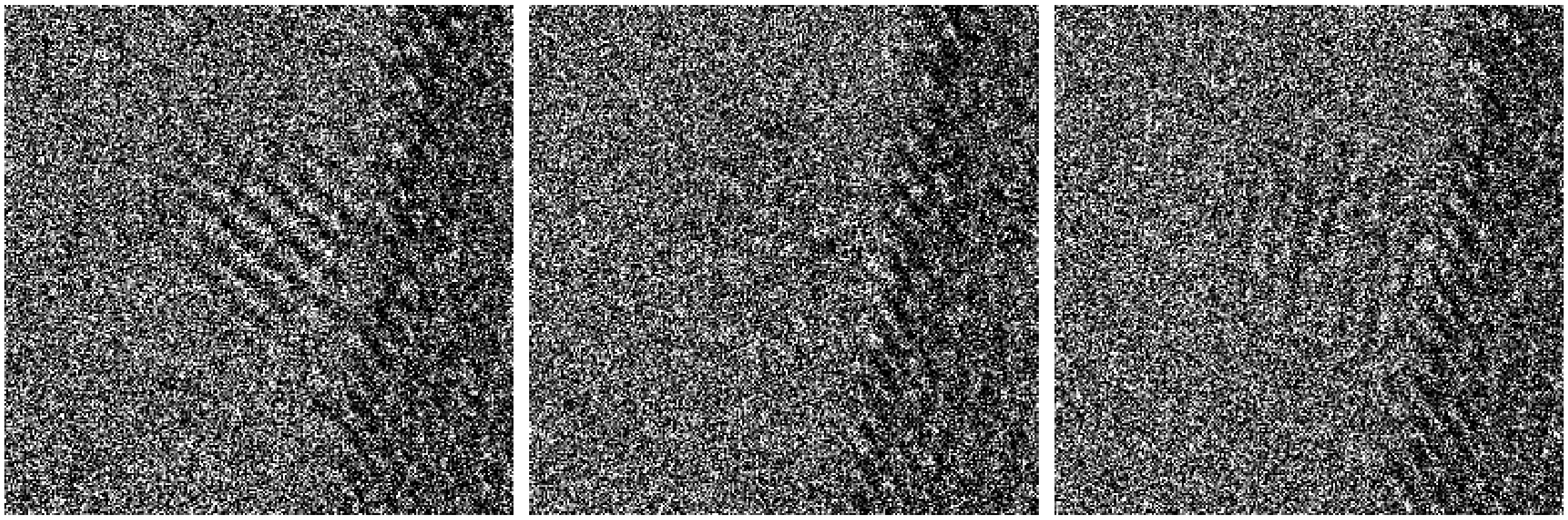

***Figure S3B:*** Three frames from a Pt nanoparticle separated in time by about 1s showing fluxionality. The images were recorded at a dose rate of 200 $e^-/Å^2/s$.

## Supplement 4: Detecting Adlayers and Shearing in Raw Data

The denoiser attempts to estimate the most likely value of a pixel by learning from all the surrounding pixels in space and time. It also learns about the motifs or patterns that appear in the images from the entire movie and uses this in up sampling to create the reconstructed denoised image. When run correctly, the denoiser should not create a motive out of nothing. For a feature to be real in the denoised output, there should be evidence for the motif in the raw data. The observations in the denoiser output of adlayers on the surface and smeared atomic columns during stacking fault formation are scientific important. To validate this observation we demonstrate that, when you know when and where to look, such features can be observed in the raw data.

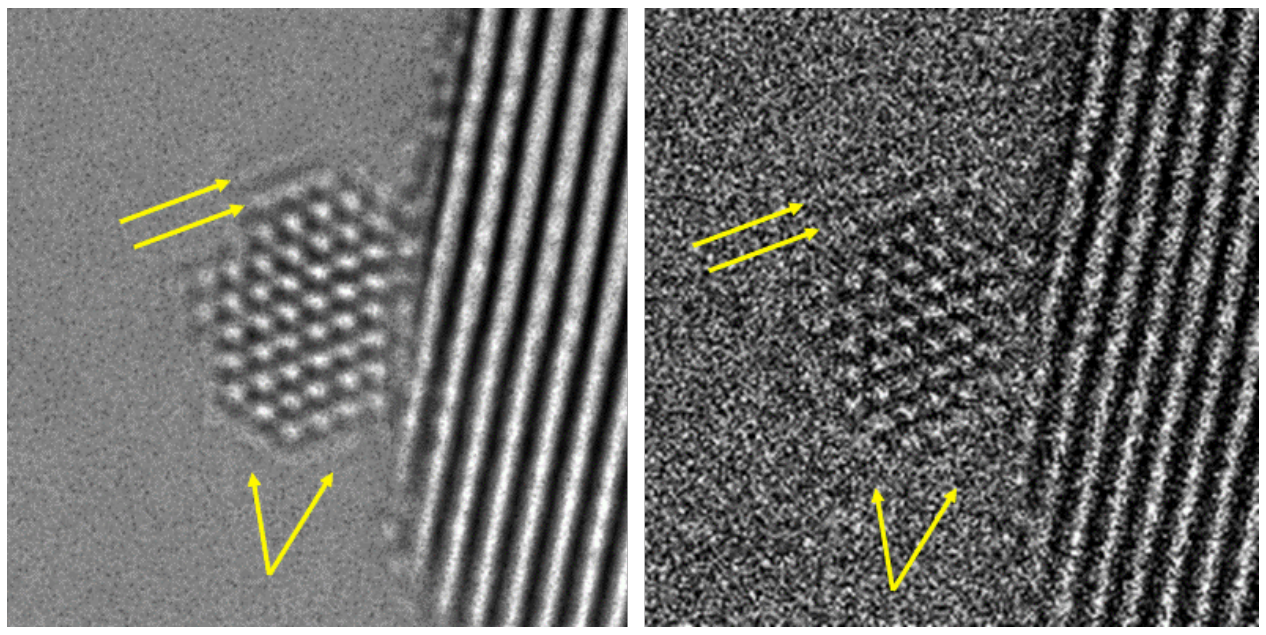

**Figure S4A:** Left - Denoise image showing 4 diffuse adlayers. Right - The raw data showing the same area and time after 3 x 3 binning and 4 frame sum. The raw image is still noisy but diffuse bright surface lines are clearly differentiated from the Pt atomic columns in the subsurface region. This showing that the adlayers are present in the raw data.

To demonstrate this, we first look at the denoised movie and locate frames that show relatively long-lived adlayer features on the surface that are present for up to 4 frames (50 ms) and locate the corresponding frames of the raw data. To improve the signal-to-noise in the raw data we perform 3 x 3 bin and 4 frame summing and frame summing to increase the signal-to-noise (at the expense of spatiotemporal resolution). This improves the vacuum SNR from 0.45 to 2.7 making it easier to see structural detail in the image. **Figure S4A** shows a denoised image from a frame that showing 4 diffuse adlayers. The raw data shows the same area and time after 3 x 3 binning and 4 frame summing. Though the raw image is still noisy, it clearly shows diffuse bright lines at the surface which are clearly differentiated from the Pt atomic columns in the subsurface region. This shows that the adlayers are present in the raw data. They are not an artifact of denoising.

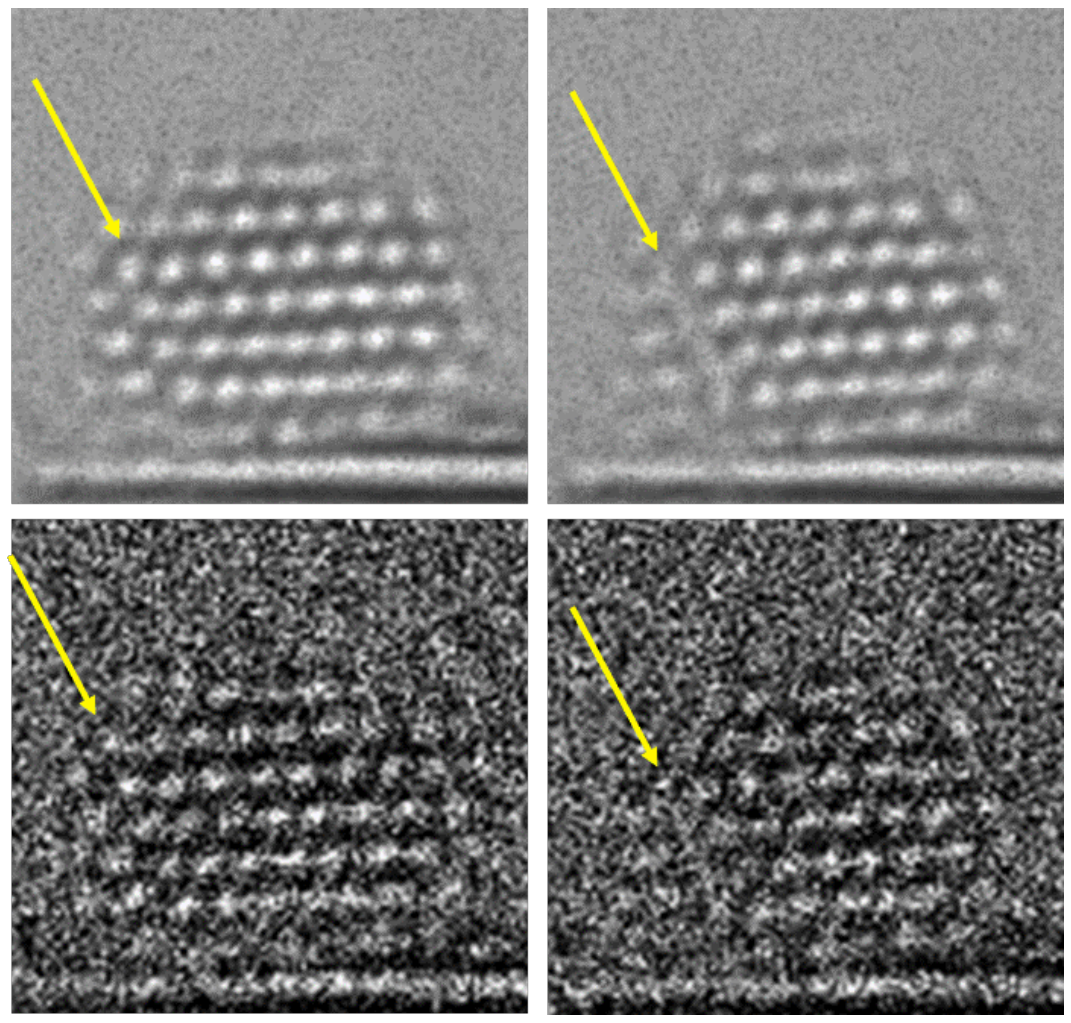

**Figure S4B:** Upper **-** Denoised image showing frames 26 ms before (left) and during shearing (right) to form a stacking fault. The yellow arrow shows the shearing plane which appears as a line when the stacking fault is forming. Lower - the raw data shows the same area and time after 3 x 3 binning and 4 frame sum. The image is noisy but before shearing (left) Pt atomic columns are clearly observed and during shearing (right) the row of atomic columns show as a nearly continuous line. This is in agreement with the denoiser output.

**Figure S4B** shows a pair of denoised images recorded 26 ms before and during a shearing event that creates a stacking fault. During the shearing event, the atomic columns are streaked giving rise to a bright line.  The raw data showing the same area and time after 3 x 3 binning and 4 frame sum show identical features before and during the shearing.

These two validation tests show that the denoiser is providing dynamic structure information that is mostly true. By significantly reducing the noise in the image, it greatly enhances our ability to observe the structural dynamics taking place on or near the nanoparticle surfaces.

Sections of the movies corresponding to Figures 2 and 3 in the main text (both raw data and denoised) are given in the attached mp4 files.

# Supplement 5: Image Simulation on the Stacking Fault/Shearing Event

Models with stacking fault and without stacking fault are generated to illustrate the shearing event. To closely match with the Pt nanoparticle structure in the experimental images, we have generated a Pt model with Winterbottom structure having total of 332 atoms using Cystal Maker software. The thickness of the atomic columns varies from 3 atoms to 17 atoms depending upon the location of the nanoparticle with surface having a smaller number of atoms in a column. For the model with shearing event (***Figure S5c***), the planes showing the stacking fault is indicated with blue arrows. A change in stacking sequence from 'ABCA**BCAB**' to 'ABCA**BCBC**' is observed. The TEM image simulation is performed using Dr. Probe software with multislice technique. The simulation pixel size is set to identical to match the experimental images and the slice thickness is 0.4 Å per slice. The parameters are defocus is 6 nm, $C_s$ value is -9 um, and $C_5$ value is 5 mm. The red dashed line in the simulated image is for visual guidance, where the red arrow corresponds to the A layer location and yellow arrow indicates the shifted position of atomic column after the shearing. The blue arrow indicated the shearing plane.

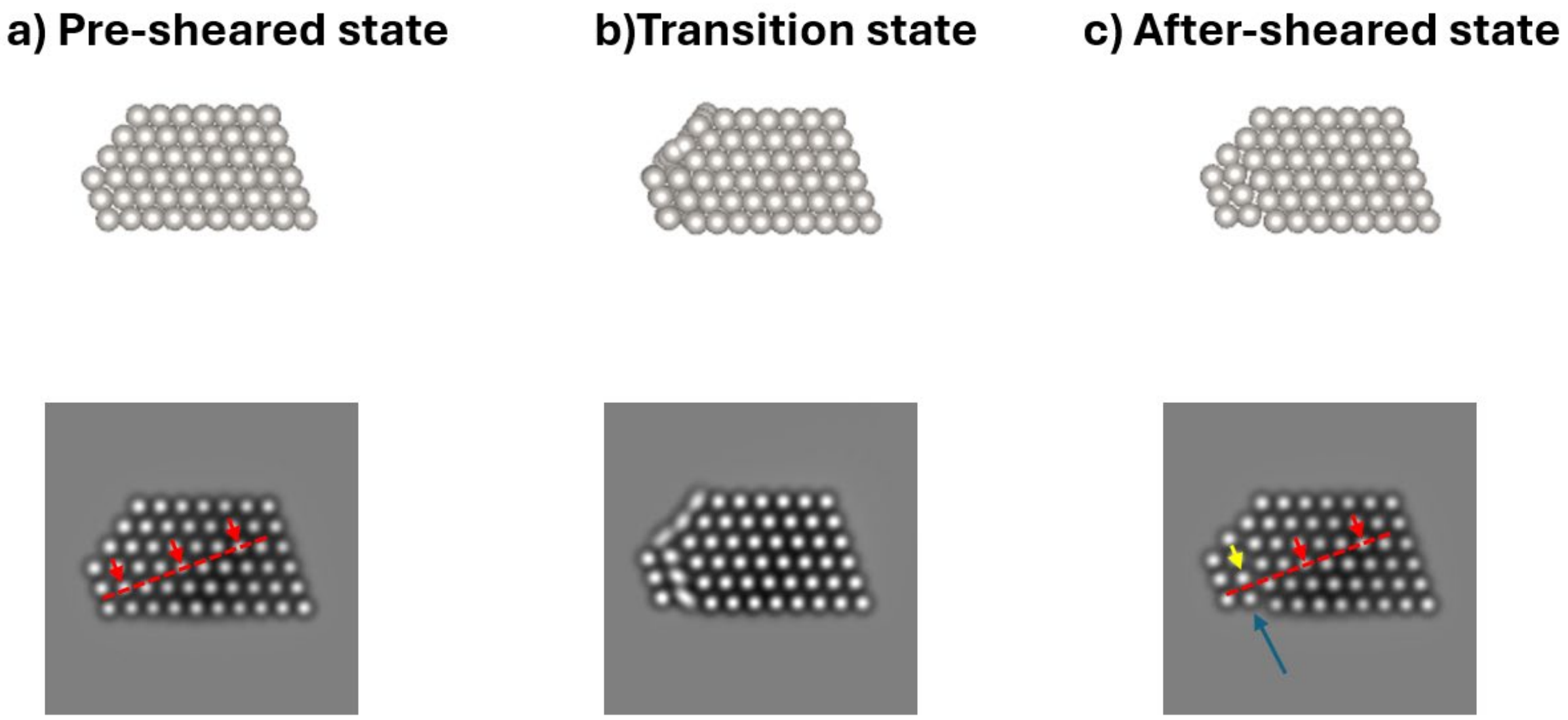

***Figure S5***: a) model and image simulation of perfect Pt nanoparticle without stacking fault at pre-shearing state b) model and image simulation of Pt nanoparticle at transition state with tilted columns and formation of the stacking fault c) model and image simulation of Pt nanoparticle after shearing. Red arrows indicate the A layer location and yellow arrow indicates the shifted position after the shearing. Blue arrow pointed at the sheared plane.

# Supplement 6: Persistent homology

To detail the persistent homology that we use to analyze our images, we must first introduce cubical homology. To first calculate cubical homology, we represent an image as built up of points, lines and squares, rather than triangulated (as one would do to calculate simplicial homology). The 0-dimensional cubical homology of a binary image (which is all we consider here) is calculated by examining the degree of connectivity amongst the black regions of an image. A black pixel is said to be connected to another black pixel if it is among its 8 immediate neighbors on the grid defined by the image. More details on cubical homology can be found in Kaczynski et. al (2004)[1].

Owing to the fact that thresholding images—especially ones that may still contain some degree of noise—is not a simple task, we can instead calculate cubical homology for *all* of the binary images that are produced from thresholding the greyscale image at the unique pixel intensities of the image. For example, if $I$ is our original greyscale image and the threshold is $t$, then we calculate a binary image $I_t$, derived from converting all pixels with intensities at most $t$ to black pixels and pixels with intensities greater than $t$ to white pixels. This produces a sequence of images, where the black pixels of $I_s$ are contained in larger black regions of $I_t$, if $s \leq t$.

For simplicity, we will assume all pixels in our greyscale image $I$ have a unique intensity value. Thus, if a pixel $p$ is a local minimum, then it will appear at the threshold $b(p)$ and not be surrounded by any other black pixels. The pixel $p$ creates or gives birth to a connected component $C_p$. As the threshold $t$ increases, $C_p$ will continue to gain more and more black pixels until it merges with another connected component $C_q$ at threshold $d$. If the threshold $b(q)$ at which $C_q$ appears is less than $b(p)$, then we say that $C_p$ dies at threshold $d(p) = d$, and associate the values $((b(p), d(p))$ to the connected component $C_p$. Thus, for set $M$ of all the local minima in the image $I$ we get a collection of points

$$PD_0 = \{(b(p), d(p)) : p \in M\}$$

called a *persistence diagram*. Note that the subscript 0 is present because we could also consider the 1-dimensional homology, which basically considers the white regions of the image enclosed by black pixels. Owing to the duality between 0- and 1-dimensional persistent homology in images[2], we do not pursue this here. A thorough introduction to persistent homology in the context of data science can be seen in Carlsson and Vejdemo-Johansson (2021)[3] and a more detailed treatment of these topological methods seen in this supplement can be found in Thomas et. al (2023)[4].

The persistence diagram $PD_0$ can be plotted in the plane, as is seen in Figure S6.1 for the two denoised nanoparticle images in ***Figure S6.2.*** To summarize the shape content present in the image via a single numerical value, we chose the ALPS statistic, which was devised in Thomas et. al (2023)[4]. The ALPS statistic is desirable because it yields conclusions in a statistical hypothesis test that are very close to those derived using the estimated number of atomic columns within the image (ibid.). However, the ALPS statistic requires no thresholding or tuning. The specific hypothesis test in question was whether the image represented pure noise or some nanoparticle signal.

The ALPS statistic can be equivalently defined as the area under a curve (cf. ***Figure S6.3***) as well as a weighted sum of the *persistence lifetimes*

$$l(p) = d(p) - b(p)$$

in $PD_0$ (see Thomas et al. 2023, Proposition 5.1). It can capture the degree of structure present in a way that standard Euclidean summaries cannot and, as mentioned, has no parameters to tune. For example, in ***Figure S6.4***, we see images with identical pixel intensity distributions. However, their ALPS statistics[*] show an increase in order. As this order was induced by increasing the scale parameter in a Gaussian filter, this seems to also correspond to a "zooming in" effect. Furthermore, in the calculation of the persistence diagram $PD_0$ of an image $I$—as the means of calculating the ALPS statistic—we ascertain the location of local minima $M$ of an image along with the persistence lifetimes $l(p)$ of those pixels. These values can be conveniently plotted on the original/smoothed image—see the third column of Figure S6.2.

[*] Upon smoothing the images in Figure S6.1. We also refer to this as the smoothed ALPS statistic.

For our analyses, we used the detectda[5] Python package. This package was specifically designed to allow for the easy calculation of nanoparticle dynamics via cubical persistent homology. The package implements the **detectda** algorithm, which upon specification of polygonal region (see ***Figure S6.5***), yields topological properties of images, within the restricted region, such as the ALPS statistic, persistent entropy[6,7], and many more. An example of the **detectda** algorithm can be seen in Figure S6.2 below. A crucial component of the **detectda** algorithm is to first convolve the image with some filter—here taken to be a Gaussian filter applied in the same manner as Thomas et. al[4]. The Gaussian filter was justified by its prior use in the literature and its ability to preserve local minima within a discrete signal to a high degree. We chose a smoothing parameter of $\sigma = 2$, for our symmetric Gaussian kernel, which yielded the best results in this study as well as our prior study[4].

Another desirable property of the ALPS statistics is that small perturbations of an image—noisy or otherwise—do not induce large changes in the (smoothed) ALPS statistic. This follows from Proposition 5 in Solomon and Bendich (2024)[8], which shows that the Bottleneck distance (for example) between the persistence diagrams is bounded by the supremum distance between the pixel intensities of the raw images. Furthermore, it can be shown the distance between ALPS statistics for two persistence diagrams is bounded by the bottleneck distances of the respective diagrams. This can be done similar to how it is accomplished for the persistent entropy[6], but we omit the proof here.

The smoothed ALPS statistic contrasts with the ALPS statistic taken from persistence diagram calculated from the raw, unfiltered image, but it is the one we use exclusively in this study, so we will often drop the qualifier “smoothed”.

The ALPS statistic as introduced above is sensitive to the size of the imaging region. However, it has been established by one of the authors in a paper in preparation (Thomas (2024), Convergence of Persistence Diagrams for Discrete Time Stationary Processes) that for a reasonably behaved one-dimensional stationary discrete signal of length *n* with a sufficiently large number of points, that

$$\text{ALPS} \approx \beta_0 + \beta_1 * \ln(n)$$

for some constants $\beta_0$ and $\beta_1$. As the proof therein requires no special techniques relating to the dimensionality of the signal, this property should also hold for two-dimensional signals as well. Suppose our image has a *n* pixels. If this image was no different from noise, its smoothed ALPS statistic would be similar to that of the vacuum region within an image. Smoothed i.i.d. noise is ergodic so satisfies the assumption required of the one-dimensional signal and yields for vacuum regions that

$$\frac{\text{ALPS} - \beta_0}{\beta_1 * \ln(n)} \approx 1$$

By looking at the smoothed ALPS statistic for particles of various sizes, we may estimate $\beta_0$ and $\beta_1$, and denote these estimated values $\widehat{\beta_0}$ and $\widehat{\beta_1}$. $\widehat{\beta_0}$ is equal to -5.08 and $\widehat{\beta_1}$ is equal to 0.705 for the specific data shown in the main documents. Those constants are generated by measuring regions of different size in the vacuum region of the experimental image. Figure S6.6 shows the linear relationship between the natural logarithm of the number of total pixels in the region of interest versus the ALPS statistic. Thus, we arrive at the quantity we deem in the supplement as the *standardized ALPS statistic:*

$$\text{ALPS}^* = \frac{\text{ALPS} - \widehat{\beta_0}}{\widehat{\beta_1} * \ln(n)} = \frac{\text{ALPS} + 5.08}{0.705 * \ln(n)}$$

However, throughout the main document we only employ the standardized ALPS$^*$, so we simply denote it as ALPS. As a result of the above, an ALPS$^*$ value around 1 means that a given image is topologically indistinguishable from noise, at least with respect to the ALPS statistic.

We see the ALPS$^*$ values for some images in Figure 4 of the main document in ***Figure S6.5.*** From this figure, there is an evident transition as the ALPS$^*$ statistic goes from 1 to the maximum value of 1.563. The persistence diagrams for the images with the minimum and maximum ALPS$^*$ values for the nanoparticle video of ***Figure 4*** in the main document can be seen in ***Figure S6.1,*** with the persistence lifetimes marked as the distances from the diagonal. It is evident from ***Figure S6.1*** and ***Figure S6.3*** that frame 754 contains many more short-lived "shallower" features and frame 41 contains features that live on much longer.

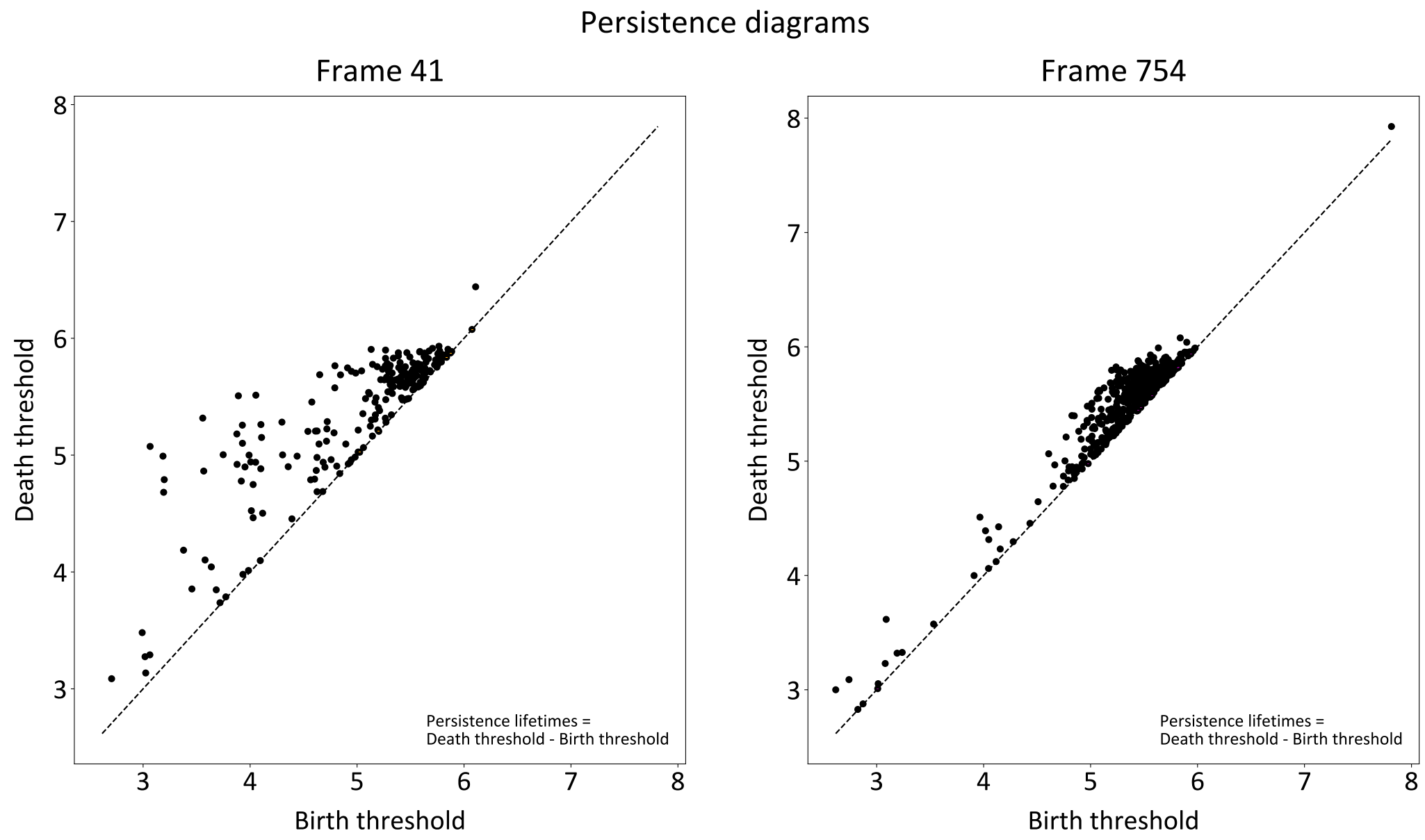

***Figure S6.1***: Persistence diagrams for the images in frames 41 and 754 of the image series seen in Figure 4 of the main article. A clear difference between the plots is the presence of connected components in the persistence diagram in frame 41 is the existence of much longer-lived connected components. This corresponds to atomic structure present in the image.

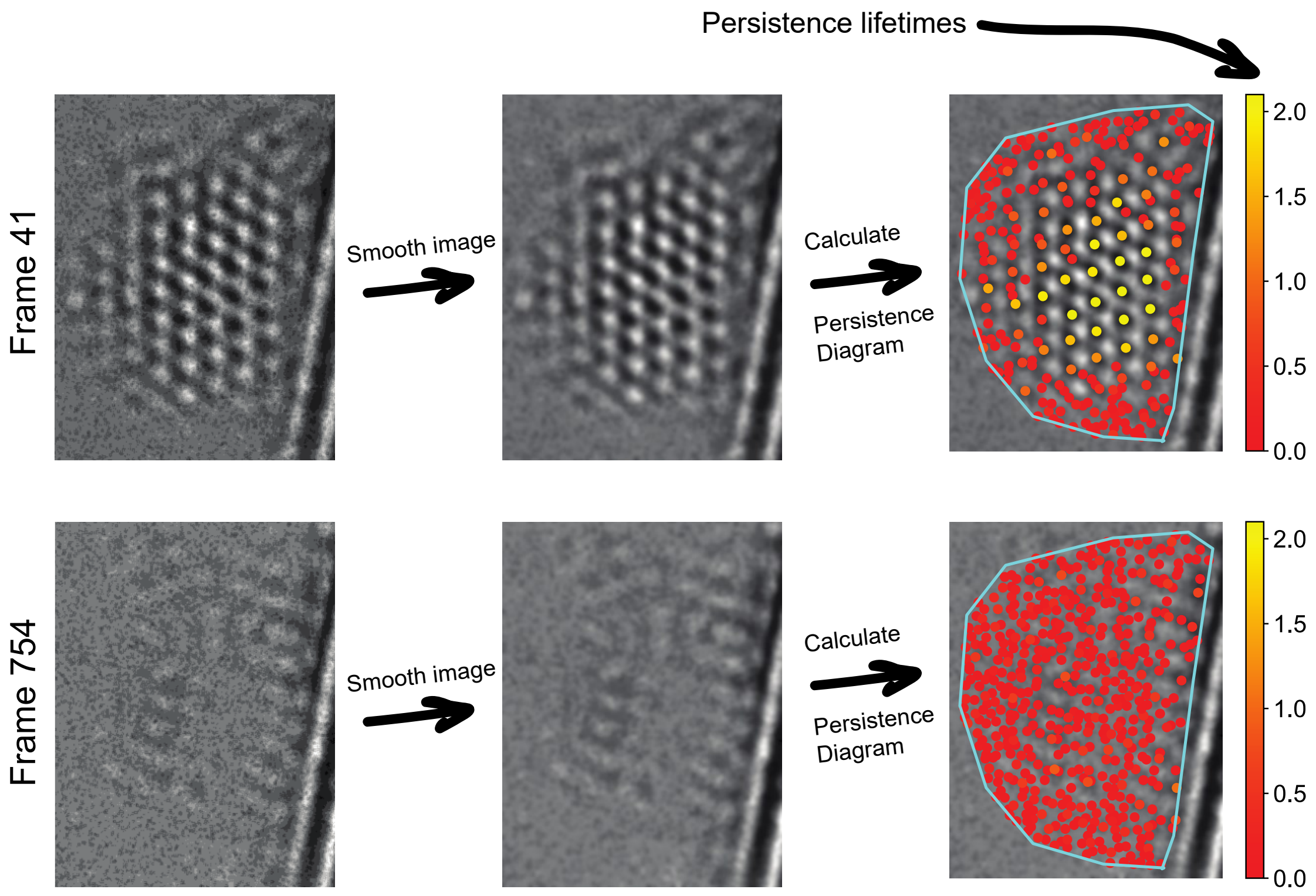

***Figure S6.2***: Nanoparticle processing pipeline for frames 41 and 754 from the image series in Figure 4 of the main article. We take the denoised image (processed according to our unsupervised denoiser), smooth said image with a Gaussian filter with $\sigma = 2$, and then apply the **detectda** algorithm to it, to determine the location and the persistence lifetimes within the cyan polygon for each nanoparticle. This information is derived from the persistence diagrams and is used to calculate both the unstandardized and standardized ALPS statistic.

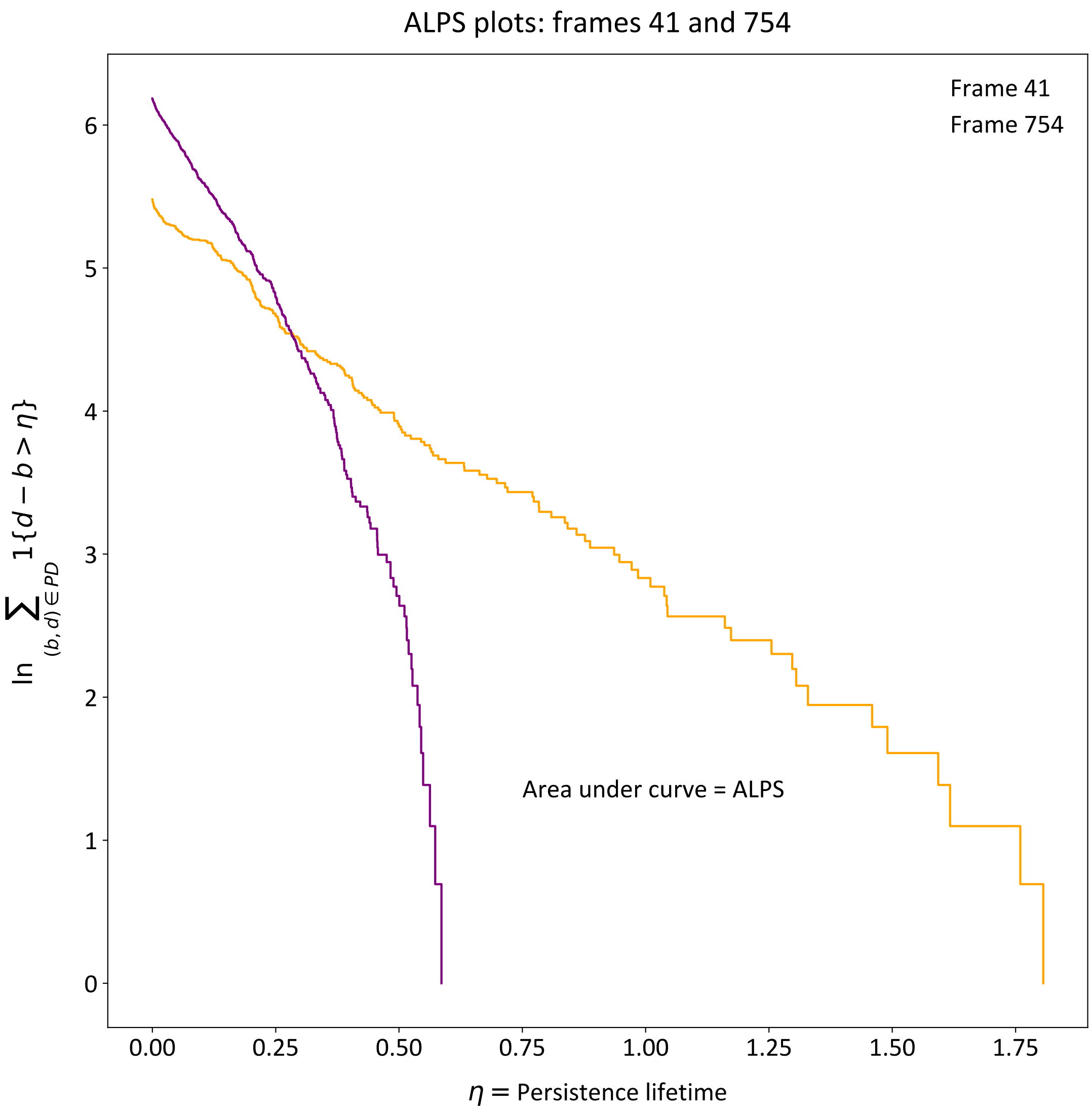

***Figure S6.3:*** Plot of curves that yield the unstandardized ALPS statistic for images in frames 41 and 754 of the image series in Figure 4 of the main article. One can see that the noise is frame 41 is much more prominent (and numerous) than that of frame 754.

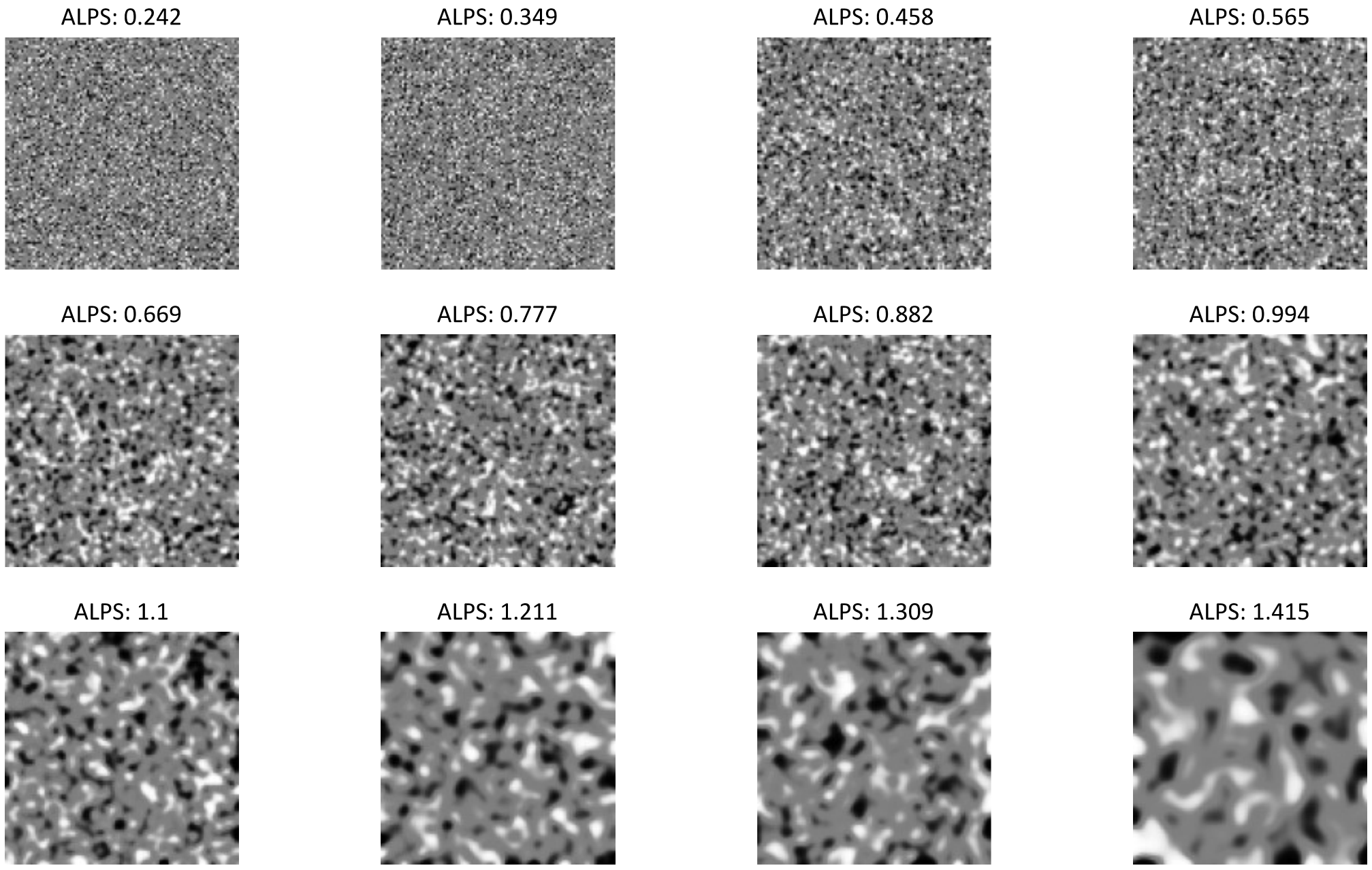

***Figure S6.4:*** Unstandardized (smoothed) ALPS statistics for 12 images with identical pixel intensity distributions. An increase of the ALPS statistic roughly corresponds to an increase in order/scale. Images were filtered according to a symmetric Gaussian kernel with $\sigma = 2$, prior to calculating the ALPS statistic. Unfiltered images are displayed.

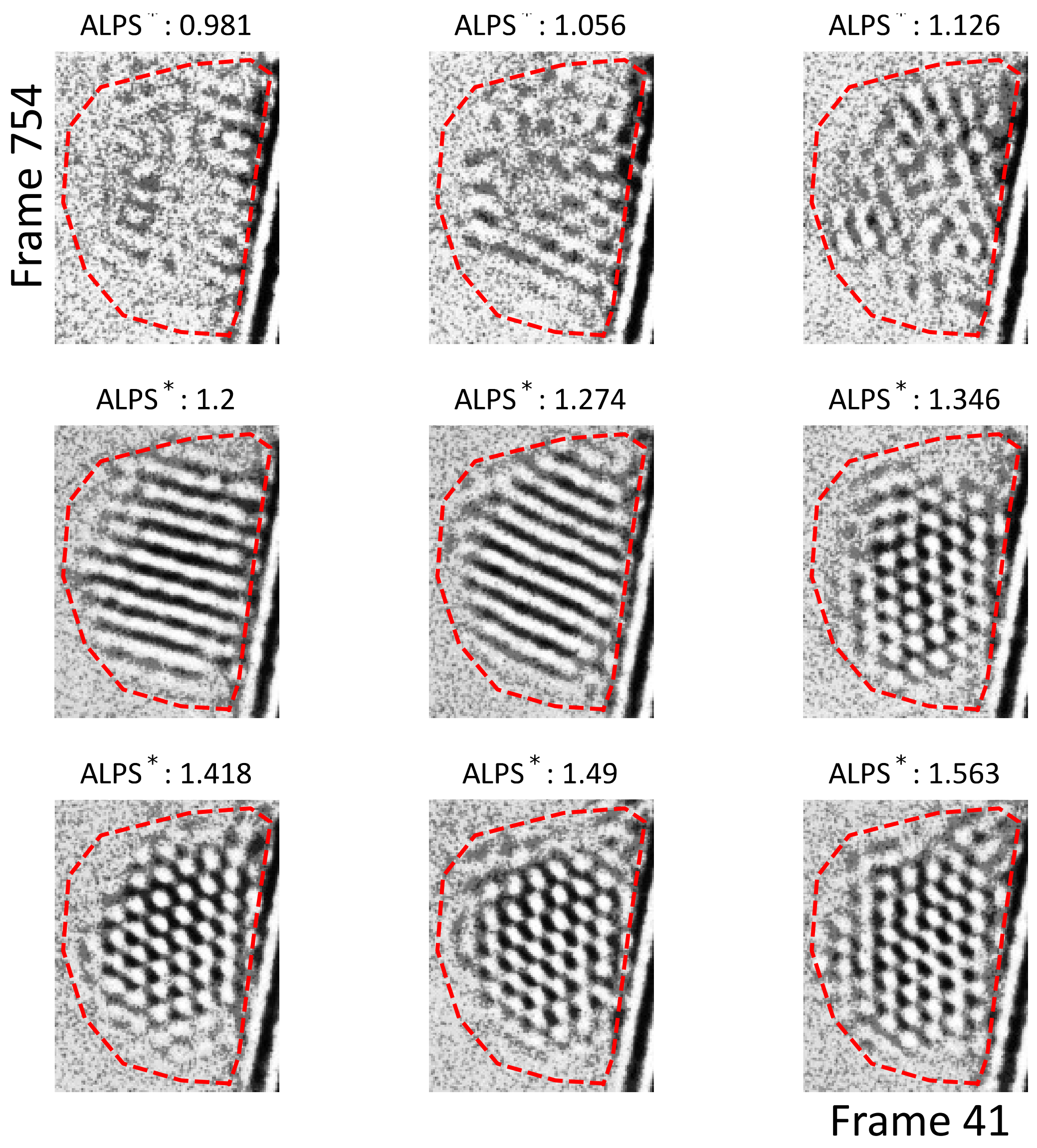

***Figure S6.5:*** Standardized ALPS statistics for 9 images from Figure 4 in the main document. Images were filtered according to a symmetric Gaussian kernel with $\sigma = 2$, prior to calculating the ALPS statistic. Images have been histogram equalized after calculating the ALPS statistic to facilitate comparisons. Only points in the persistence diagram corresponding to pixels located in the cyan (dashed) polygon were used in calculating the ALPS statistic.

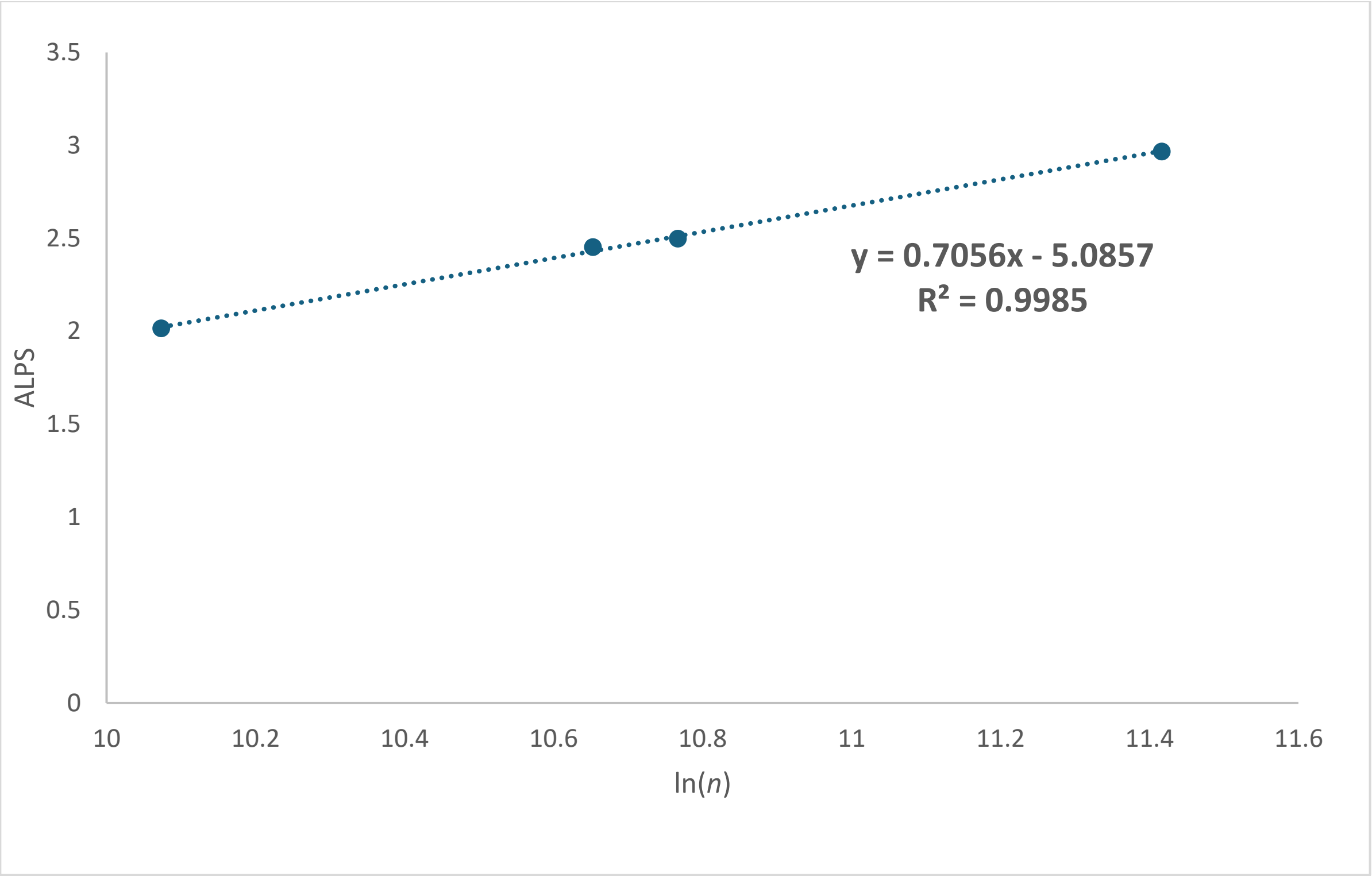

***Figure S6.6:*** Plot of ALPS statistics for regions of different size in the vacuum. The dotted blue lines are the linear best-fit and the linear equation is displayed. The intercept and slope are corresponding to $\widehat{\beta_0}$ and $\widehat{\beta_1}$.